\documentclass[english,draftclsnofoot,one column, 12pt]{IEEEtran}
\usepackage[T1]{fontenc}
\usepackage[latin9]{inputenc}
\usepackage{verbatim}
\usepackage{bm}
\usepackage{amsmath}
\usepackage{graphicx}
\usepackage{amssymb}

\makeatletter

\providecommand{\tabularnewline}{\\}

\IEEEoverridecommandlockouts

\usepackage{amsfonts}\usepackage{mathrsfs}\usepackage{bm}\usepackage{algorithmicx}
\usepackage{array}\usepackage{mdwmath}\usepackage{mdwtab}\usepackage[caption=false,font=footnotesize]{subfig}\usepackage{fixltx2e}
\usepackage{footnote}\usepackage{tabularx}\usepackage{multirow}\usepackage{cite}
\usepackage{fixltx2e}\usepackage{braket}\usepackage{algorithmicx}\usepackage{algpseudocode}\usepackage{multirow}

\usepackage{url}
\makeatletter

\newcommand{\Rmnum}[1]{\expandafter\@slowromancap\romannumeral #1@}
\makeatother

\makeatother

\usepackage{babel}

\makeatother

\usepackage{babel}

\begin{document}

\title{ Distributed Interference Management Policies
for Heterogeneous Small Cell Networks}

\author{\IEEEauthorblockN{{ Kartik Ahuja, Yuanzhang Xiao and
Mihaela van der Schaar }}\\
{ \IEEEauthorblockA{Department of Electrical Engineering,
UCLA, Los Angeles, CA, 90095}\\
{Email: ahujak@ucla.edu, yxiao@seas.ucla.edu and
mihaela@ee.ucla.edu}} 
\vspace{-2em}\thispagestyle{plain}\vspace{-1.35em} \pagestyle{plain}}
\maketitle
\begin{abstract}
We study the problem of ${\it \textit{distributed}}$ interference
management in a network of heterogeneous small cells with different
cell sizes, different numbers of user equipments (UEs) served, and
different throughput requirements by UEs. We consider the uplink transmission,
where each UE determines when and at what power level it should transmit
to its serving  small cell base station (SBS). We propose a general
framework for designing distributed interference management policies,
which exploits weak interference among non-neighboring UEs by letting
them transmit simultaneously (i.e., spatial reuse), while eliminating
strong interference among neighboring UEs by letting them transmit
in different time slots. %
{}The design of optimal interference management policies has two key
steps. Ideally, we need to find all the subsets of non-interfering
UEs, i.e., the maximal independent sets (MISs) of the interference
graph, but this is NP-hard (non-deterministic polynomial time) even
when solved in a centralized manner. Then, in order to maximize some
given network performance criterion subject to UEs' minimum throughput
requirements, we need to determine the optimal fraction of time occupied
by each MIS, which requires global information (e.g., all the UEs'
throughput requirements and channel gains). In our framework, we first
propose a $\emph{\emph{\text{\textit{distributed}}}}$ algorithm for
the UE-SBS pairs to find ${\it \textit{a subset of}}$ MISs in logarithmic
time (with respect to the number of UEs). Then we propose a novel
problem reformulation which enables UE-SBS pairs to determine the
optimal fraction of time occupied by each MIS with only local message
exchange among the neighbors in the interference graph. Despite the
fact that our interference management policies are distributed and
utilize only local information, we can $\textit{analytically}$ bound
their performance under a wide range of heterogeneous deployment scenarios
in terms of the competitive ratio with respect to the optimal network
performance, which can only be obtained in a centralized manner with
NP complexity. Remarkably, we prove that the competitive ratio is
independent of the network size. Through extensive simulations, we
show that our proposed policies achieve significant performance improvements
(ranging from 150\% to 700\%) over state-of-the-art policies.

\vspace{-1.7em}
\end{abstract}

\section{Introduction}

Dense deployment of low-cost heterogeneous small cells (e.g. picocells,
femtocells) has become one of the most effective solutions to accommodate
the exploding demand for wireless spectrum \cite{HossainLeNiyato}\cite{ghosh2012heterogeneous}\cite{andrews2012femtocells}.
On one hand, dense deployment of small cells significantly shortens
the distances between small cell base stations (SBSs) and their corresponding
user equipments (UEs), thereby boosting the network capacity. On the
other hand, dense deployment also shortens the distances between neighboring
SBSs, thereby potentially increasing the inter-cell interference.
Hence, while the solution provided by the dense deployment of small
cells is promising, its success depends crucially on interference
management by the small cells. Efficient interference management is
even more challenging in heterogeneous small cell networks, due to
the lack of central coordinators, compared to that in traditional
cellular networks.

In this paper, we propose a novel framework for designing interference
management policies in the uplink of small cell networks, which specify
when and at what power level each UE should transmit%
\footnote{Although we focus on uplink transmissions in this paper, our framework
can be easily applied to downlink transmissions.%
}. Our proposed design framework and the resulting interference management
policies fulfill $\emph{all}$ the following important requirements: 
\begin{itemize}
\item ${\it \textit{Deployment of heterogeneous small cell networks}}$:
Existing deployments of small cell networks exhibit significant heterogeneity
such as different types of small cells (picocells and femtocells),
different cell sizes, different number of UEs served, different UEs'
throughput requirements etc. 
\item ${\it \textit{Interference avoidance and spatial reuse}}$: Effective
interference management policies should take into account the strong
interference among neighboring UEs, as well as the weak interference
among non-neighboring UEs. Hence, the policies should effectively
avoid interference among neighboring UEs and use spatial reuse to
take advantage of the weak interference among non-neighboring UEs.
\item $\textit{Distributed implementation with local information and message exchange}$:
Since there is no central coordinator in small cell networks, interference
management policies need to be computed and implemented by the UEs
in a distributed manner, by exchanging only local information through
local message exchanges among neighboring UE-SBS pairs.
\item ${\it \textit{Scalability to large networks}}$: Small cells are often
deployed over a large scale (e.g., in a city). Effective interference
management policies should scale in large networks, namely achieve
efficient network performance while maintaining low computational
complexity.
\item ${\it \textit{Ability to optimize different network performance criteria}}$:
Under different deployment scenarios the small cell networks may have
different performance criteria, e.g., weighted sum throughput or max-min
fairness. The design framework should be general and should prescribe
different policies to optimize different network performance criteria.
\item ${\it \textit{Performance guarantees for individual UEs}}$: Effective
interference management should provide performance guarantees (e.g.,
minimum throughput guarantees) for individual UEs.
\end{itemize}
As we will discuss in detail in Section \ref{sec:related works},
existing state-of-the-art policies for interference management cannot
simultaneously fulfill all of the above requirements.

Next, we describe our key results and major contributions:

1. We propose a general framework for designing distributed interference
management policies that maximizes the given network performance criterion
subject to each UE's minimum throughput requirements. The proposed
policies schedule maximal independent sets (MISs)%
\footnote{Consider the interference graph of the network, where each vertex
is a UE-SBS pair and each edge indicates strong interference between
the two vertices. An independent set (IS) is a set of vertices in
which no pair is connected by an edge. An IS is a MIS if it is not
a proper subset of another IS. %
} of the interference graph to transmit in each time slot. In this
way, they avoid strong interference among neighboring UEs (since neighboring
UEs cannot be in the same MIS), and efficiently exploit the weak interference
among UEs in a MIS by letting them to transmit at the same time.

2. We propose a distributed algorithm for the UEs to determine a subset
of MISs. The subset of MISs generated ensures that each UE belongs
to at least one MIS in this subset. Moreover, the subset of MISs can
be generated in a distributed manner in logarithmic time (logarithmic
in the number of UEs in the network) for bounded-degree interference
graphs%
\footnote{Bounded-degree graphs are the graphs whose maximum degree can be bounded
by a constant independent of the size of the graph, i.e., $\Delta=\mathcal{O}(1)$.
As we will show in Theorem 5, for the interference graphs that are
not bounded-degree graphs, even the centralized solution, given all
the MISs, cannot satisfy the minimum throughput requirements.%
}. The logarithmic convergence time is significantly faster than the
time (linear or quadratic in the number of UEs) required by the distributed
algorithms for generating subsets of MISs in \cite{ramaswami1989distributed,cidon1989distributed,ephremides1990scheduling}.

3. Given the computed subsets of MISs, we propose a distributed algorithm
in which each UE determines the optimal fractions of time occupied
by the MISs with only local message exchange. The message is exchanged
only among the UE-SBS pairs that strongly interfere with each other,
i.e. among neighbors in the interference graph. The distributed algorithm
will output the optimal fractions of time for each MIS such that the
given network performance criterion is maximized subject to the minimum
throughput requirements.

4. Under a wide range of conditions, we analytically characterize
the competitive ratio of the proposed distributed policy with respect
to the optimal network performance. Importantly, we prove that the
competitive ratio is independent of the network size, which demonstrates
the scalability of our proposed policy in large networks. Remarkably,
the constant competitive ratio is achieved even though our proposed
policy requires only local information, is distributed, and can be
computed fast, while the optimal network performance can only be obtained
in a centralized manner with global information (e.g., all the UEs'
channel gains, maximum transmit power levels, minimum throughput requirements)
and NP (non-deterministic polynomial time) complexity. 

5. Through simulations, we demonstrate significant (from 160\% to
700 \%) performance gains over state-of-the-art policies. Moreover,
we show that our proposed policies can be easily adapted to a variety
of heterogeneous deployment scenarios, with dynamic entry and exit
of UEs.


The rest of the paper is organized as follows. In Section \ref{sec:related works}
we discuss the related works and their limitations. We describe the
system model in Section \ref{sec:System-Model-l}. Then we formulate
the interference management problem and give a motivating example
in Section \ref{sec:Problem-Formulation-and}. We propose the design
framework in Section \ref{sec:The-Distributed-Design}, and demonstrate
the performance gain of our proposed policies in Section \ref{sec:Illustrative-Results}.
Finally, we conclude the paper in Section \ref{sec:Conclusion}.

\section{Related Works}

\label{sec:related works} 

State-of-the-art interference management policies can be divided into
three main categories: policies based on power control, policies based
on spatial reuse, and policies based on joint spatial reuse and power
control. 

\vspace{-0.8em}

\subsection{Distributed Interference Management Based on Power Control}

Policies based on distributed power control, with representative references
\cite{huang2006distributed,saraydar2001pricing,saraydar2002efficient,Decen-pow-game,dist-pow-ofdma,dist-res-femto,jointerf,hv-zhu}
have been used for interference management in both cellular and ad-hoc
networks. In these policies, all the UEs in the network transmit at
a $\emph{constant}$ power all the time (provided that the system
parameters remain the same)%
\footnote{Although some power control policies \cite{huang2006distributed,saraydar2001pricing,Decen-pow-game}
go through a transient period of adjusting the power levels before
the convergence to the optimal power levels, the users maintain constant
power levels after the convergence.%
}. The major limitation of policies based on power control is the difficulty
in providing minimum throughput guarantees for each UE, especially
in the presence of strong interference. Some works \cite{huang2006distributed,saraydar2001pricing,Decen-pow-game}
use pricing to mitigate the strong interference. However, they \cite{huang2006distributed,saraydar2001pricing,Decen-pow-game}
cannot strictly guarantee the UEs' minimum throughput requirements.
Indeed, the low throughput experienced by some users, caused by strong
interference, is the fundamental limitation of such power control
approaches - even the optimal power control policy obtained by a central
controller \cite{chiang2007power,gjendemsj2008binary} can be inefficient
\footnote{In the case of average sum throughput maximization given the minimum
average throughput constraints of the UEs, the power control policies
are inefficient if the feasible rate region is non-convex \cite{stanczak2009fundamentals}
.%
}. Since strong interference is very common in dense small cell deployments
(e.g. in offices and apartments where SBSs are installed close to
each other \cite{res-dense-dep}), more efficient policies are required
which can guarantee the individual UEs' throughput requirements. Also,
there exist a different strand of work based on \cite{foschini1993simple}
which proposes a distributed algorithm to achieve the desired minimum
throughput requirement for each UE. However, these works cannot optimize
network performance criterion such as weighted sum throughput, max-min
fairness etc. and hence are suboptimal.

\vspace{-1em}

\subsection{Distributed Spatial Reuse Based on Maximal Independent Sets}

An efficient solution to mitigate strong interference is spatial reuse,
in which only a subset of UEs (which do not significantly interfere
with each other) transmit at the same time. Spatial Time reuse based
Time Division Multiple Access (STDMA) has been widely used in existing
works on broadcast scheduling in multi-hop networks\cite{ramaswami1989distributed,cidon1989distributed,ephremides1990scheduling}%
\footnote{These works \cite{ramaswami1989distributed,cidon1989distributed,ephremides1990scheduling}
do not have the exactly same model as in our setting. However, these
works can be adapted to our model. Hence, we also compare with these
works to have a comprehensive literature review.%
}. Specifically, these policies construct a cyclic schedule such that
in each time slot an MIS of the interference graph is scheduled. The
constructed schedule ensures that each UE is scheduled at least once
in the cycle. %
{}

In terms of performance, STDMA policies \cite{ramaswami1989distributed,cidon1989distributed,ephremides1990scheduling}
cannot guarantee the minimum throughput requirement of each UE, and
usually adopt a fixed scheduling (i.e. follow a fixed order in which
the MISs are scheduled), which may be very inefficient depending on
the given network performance criteria. For example, the policies
in \cite{ephremides1990scheduling} are inefficient in terms of fairness.
In terms of complexity, for the distributed generation of the subsets
of MISs, the STDMA policies in \cite{ramaswami1989distributed,cidon1989distributed,ephremides1990scheduling}
require an ordering of all the UEs, and have a computational complexity
(in terms of the number of steps executed by the algorithm) that scales
as $\mathcal{O}(|V|))$ (in \cite{cidon1989distributed,ephremides1990scheduling})
or $\mathcal{O}(|V||E|))$ (in \cite{ramaswami1989distributed}),
where $|V|$ and $|E|$ are the number of vertices/UEs and the number
of edges in the interference graph, respectively. Hence, in large-scale
dense deployments, the complexity grows superlinearly with the number
of UEs, making the policies difficult to compute. By contrast, our
proposed distributed algorithm for generating subsets of MISs does
not require the ordering of all the UEs, and has a complexity that
scales as $\mathcal{O}(\log|V|)$, namely sublinearly with the number
of the UEs, for bounded-degree graphs.%
{}%
\footnote{ As will be shown in Theorem 5, for graphs which are not bounded degree
graphs, even a centralized solution based on all the MISs cannot satisfy
the minimum throughput requirements.%
}

Finally, the STDMA policies in \cite{ramaswami1989distributed,cidon1989distributed,ephremides1990scheduling}
are designed for the MAC layer and assume that all the UEs are homogeneous
at the physical layer. In practice, different UEs are heterogeneous
due to their different distances from their SBSs, their different
maximum transmit power levels, etc. This heterogeneity is important,
and will be considered in our design framework.

\vspace{-1em}

\subsection{Distributed Power Control and Spatial Reuse For Multi-Cell Networks}

As we have discussed, the works in the above two categories either
focus on distributed power control in the physical layer \cite{huang2006distributed,saraydar2001pricing,Decen-pow-game}
or focus on distributed spatial reuse in the MAC layer \cite{ramaswami1989distributed,cidon1989distributed,ephremides1990scheduling}.
Similar to our paper, some works (representative references\cite{kiani2006maximizing,kiani2007maximizing,kiani2008optimal,gesbert2007adaptation,chen2006joint}
) adopted a cross-layer approach and proposed distributed joint power
control and spatial reuse for multi-cell networks. However, although
these works schedule a subset of UEs to transmit at the same time,
the subset is not the MIS of the interference graph \cite{kiani2008optimal,gesbert2007adaptation}.
For example, the policies in \cite{kiani2008optimal,gesbert2007adaptation}
schedule one UE from each small cell at the same time, even if some
UEs are from small cells very close to each other. In this case, the
UEs will experience strong inter-cell interference. Hence, the works
in \cite{kiani2008optimal,gesbert2007adaptation} cannot perfectly
eliminate strong interference from neighboring cells and exploit weak
interference from non-neighboring cells. Moreover, the works in \cite{kiani2006maximizing,kiani2007maximizing,kiani2008optimal,gesbert2007adaptation,chen2006joint}
cannot provide minimum throughput guarantees for the UEs.

\section{System Model \label{sec:System-Model-l}}

\vspace{-0.5em}

\subsection{Heterogeneous Network of Small Cells}

We consider a heterogeneous network of $K$ small cells operating
in the same frequency band%
\footnote{Our solutions will be based on spatial time reuse assuming every UE
uses the same frequency. Our solutions can be extended to spatial
frequency reuse, where we let different MISs operate in non-overlapping
frequency bands.%
} (see Fig. 1), which represents a common deployment scenario considered
in practice \cite{Decen-pow-game}\cite{jointerf}\cite{andrews-1}.
Note that the small cells can be of different types (e.g. picocells,
femtocells, etc.) and thereby belong to different tiers in the heterogeneous
network. Each small cell $j$ has one SBS, (SBS-$j$), which serves
a set of UEs under a closed access scenario \cite{Decen-pow-game}.
Denote the set of UEs by $\mathcal{U}=\{1,...,N\}$. We write the
association of UEs to SBSs as a mapping $T:\{1,...,N\}\rightarrow\{1,..,K\}$,
where each UE-$i$ is served by SBS-$T(i)$. We focus on the uplink
transmissions; the extension to downlink transmissions is straightforward
when each SBS serves one UE at a time (e.g. TDMA among UEs connected
to the same SBS). 

Each UE-$i$ chooses its transmit power $p_{i}$ from a compact set
$\mathcal{P}_{i}\subseteq\mathbb{R}_{+}$. We assume that $0\in\mathcal{P}_{i},\;\forall i\in\{1,...,N\}$,
namely any UE can choose not to transmit. The joint power profile
of all the UEs is denoted by $\textbf{p}=(p_{1},....,p_{N})\in\mathcal{P}\triangleq\Pi_{i=1}^{N}\mathcal{P}_{i}$.
Under the joint power profile $\textbf{p}$, the signal to interference
and noise ratio (SINR) of UE-$i$'s signal, experienced at its serving
SBS-$j=T(i)$, can be calculated as $\gamma_{i}(\textbf{p})=\frac{g_{ij}p_{i}}{\sum\limits _{k=1,k\not=i}^{N}g_{kj}p_{k}+\sigma_{j}^{2}},$
where $g_{ij}$ is the channel gain from UE-$i$ to SBS-$j$, and
$\sigma_{j}^{2}$ is the noise power at SBS $j$. 
 The UEs do not cooperate to encode their signals to avoid interference,
hence, each UE-SBS  pair treats the interference from other UEs as
white noise. Hence, each UE-$i$ gets the following throughput \cite{kiani2008optimal},
$r_{i}(\textbf{p})=\log_{2}(1+\gamma_{i}(\textbf{p}))$%
\footnote{We use the Shannon capacity here. However, our analysis is general
and applies to the throughput models that consider the modulation
scheme used.%
}. 

\vspace{-1em}

\subsection{Interference Management Policies}

The system is time slotted at $t$ = 0,1,2..., and the UEs are assumed
to be synchronized as in \cite{kiani2008optimal,gesbert2007adaptation}\cite{etkin2007spectrum}\cite{wu2009repeated}..
At the beginning of each time slot $t$, each UE-$i$ decides its
transmit power $p_{i}^{t}$ and obtains a throughput of $r_{i}(\textbf{p}^{t})$.
Each UE $i$'s strategy, denoted by $\pi_{i}:\mathbb{Z}_{+}=\{0,1,..\}\rightarrow\mathcal{P}_{i}$,
is a mapping from time $t$ to a transmission power level $p_{i}\in\mathcal{P}_{i}$.
The interference management policy is then the collection of all the
UEs' strategies, denoted by $\boldsymbol{\pi}=(\pi_{1},...,\pi_{N})$.
The average throughput for UE $i$ is given as $R_{i}(\boldsymbol{\pi})=\lim_{T\rightarrow\infty}\frac{1}{T+1}\sum\limits _{t=0}^{T}r_{i}(\textbf{p}^{t}),$ where
$\textbf{p}^{t}=(\pi_{1}(t),...,\pi_{N}(t))$ is the power profile
at time $t$. We assume the channel gain to be fixed over the considered
time horizon as in\cite{kiani2008optimal} \cite{jain2005impact,graph-cluster,graph-dynamic,lee2010interference}.
However, we will illustrate in Section \ref{sec:Illustrative-Results}
that our framework performs well under dynamic channel conditions
(due to fading, time varying channel) as well.  %
{}

An interference management policy $\bm{\pi}^{const}$ is a policy
based on power control \cite{huang2006distributed,saraydar2001pricing,Decen-pow-game}
if $\bm{\pi}^{const}(t)=\textbf{p}$ for all $t$.  As we have discussed
before, our proposed policy is based on MISs of the interference graph.
The interference graph $G$ has $N$ vertices, each of which is one
of the $N$ UE-SBS pairs. There is an edge between two pairs/vertices
if their cross interference is high (rules for deciding if interference
is high will be discussed in Section V) and let there be $M$ edges
in the graph. Given an interference graph, we write $\textbf{I}=\{I_{1},...,I_{N_{MIS}}\}$
as the set of all the MISs of the interference graph. Let $\textbf{p}^{I_{j}}$
be a power profile in which the UEs in the MIS $I_{j}$ transmit at
their maximum power levels and the other UEs do not transmit, namely
$p_{k}=p_{k}^{max}\triangleq\max\mathcal{P}_{k}\;\text{if}\; k\in I_{j}$
and $p_{k}=0$ otherwise. Let $\mathcal{P}^{MIS}=\{\textbf{p}^{I_{1}},...,\textbf{p}^{I_{N_{MIS}}}\}$
be the set of all such power profiles. Then $\bm{\pi}$ is a policy
based on MIS if $\bm{\pi}(t)\in\mathcal{P}^{MIS}$ for all $t$. We
denote the set of policies based on MISs by $\Pi^{MIS}=\{\boldsymbol{\pi}:\mathbb{Z}_{+}\rightarrow\mathcal{P}^{MIS}\}$.

\vspace{-1em}%
\begin{figure}
\begin{centering}
\includegraphics[width=4.5in]{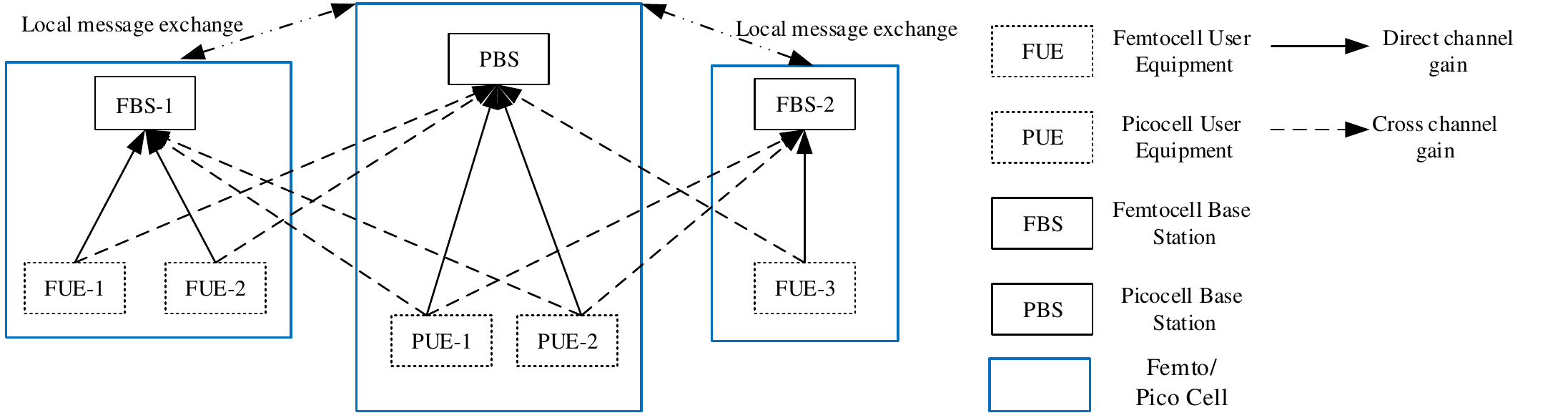}
\par\end{centering}

\caption{Illustration of a heterogeneous small cell network.}
\end{figure}

\section{Problem Formulation and A Motivating Example \label{sec:Problem-Formulation-and}}

\label{sec:prblm formuln} In this section, we formulate the interference
management policy design problem and give a motivating example to
highlight the advantages of the proposed policy over existing policies.

\vspace{-1em}

\subsection{The Interference Management Policy Design Problem}

\label{subsec:policy design} We aim to optimize a chosen network
performance criterion $W(R_{1}(\boldsymbol{\pi}),....,R_{N}(\boldsymbol{\pi}))$,
defined as a function of the UEs' average throughput. We can choose
any performance criterion that is concave in $R_{1}(\boldsymbol{\pi}),....,R_{N}(\boldsymbol{\pi})$.
For instance, $W$ can be the weighted sum of all the UEs' throughput,
i.e.$\sum\limits _{i=1}^{N}w_{i}R_{i}(\boldsymbol{\pi})$ with $\sum\limits _{i=1}^{N}w_{i}=1$
and $w_{i}\geq0$. Alternatively, the network performance can be max-min
fairness (i.e. the worst UE's throughput) and hence $W$ can be defined
as $\min_{i}R_{i}(\boldsymbol{\pi})$. The policy design problem can
be then formalized as follows:\vspace{-1.85em}

\begin{eqnarray*}
\textbf{Policy Design Problem\,(PDP)} & \max_{\bm{\pi}} & W(R_{1}(\bm{\pi}),...,R_{N}(\bm{\pi}))\\
 & \text{subject to} & R_{i}(\bm{\pi})\geq R_{i}^{min},\,\forall i\in\{1,...,N\}\end{eqnarray*}

The above design problem is very challenging to solve even in a centralized
manner (it has been shown to be NP-hard \cite{tan2011spectrum} even
when we restrict to policies based on power control $\bm{\pi}^{const}$).
Denote the optimal value of the PDP as $\text{\textbf{W}}_{\text{opt}}$.
Our goal is to develop distributed, polynomial-time algorithms to
construct policies that achieve a constant competitive ratio with
respect to $\text{\textbf{W}}_{\text{opt}}$, with the competitive
ratio independent of the network size. We achieve our goal by focusing
on policies based on MISs $\Pi^{MIS}$, among other innovations that
will be described in Section V. Next, we provide a motivating example
to demonstrate the efficiency of our proposed policy.

\section{Design Framework for Distributed Interference Management}

\subsection{Proposed Design Framework\label{sub:DesignFramework}}

\begin{figure}
\centering{}\includegraphics[width=6in]{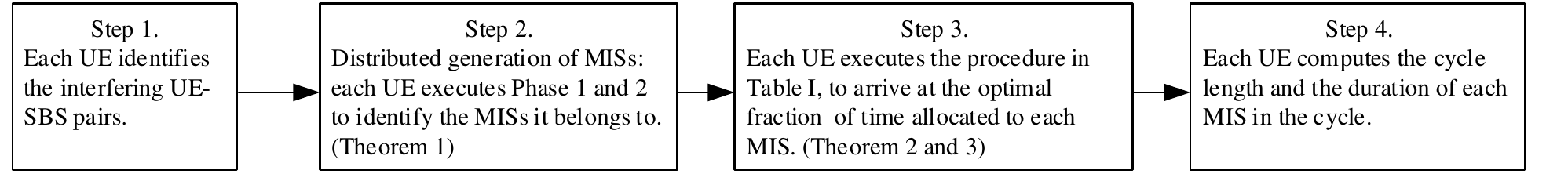}\caption{\label{fig:Steps-in-the}Steps in the Design Framework.}
\end{figure}

 Our proposed design framework (see Fig. \ref{fig:Steps-in-the})
consists of the following four steps. 

\textbf{Step 1. Identification of the interfering neighbors: }In Step
1, each UE-SBS pair identifies the UE-SBS pairs that strongly interfere
with it. Essentially, each pair obtains a $\emph{local}$ view (i.e.,
its neighbors) of the interference graph. Note that an edge exists
between two pairs if at least one of them identifies the other as
a strong interferer. 

Specifically, each UE-SBS pair is first informed of other pairs in
the geographical proximity by managing servers (e.g., femtocell controllers/gateways)
\cite{han2010automatic}\cite{lopez2011enhanced}\cite{graph-cluster}\cite{graph-dynamic}.
Then each pair can decide whether another pair is strongly interfering
based on various rules, such as rules based on Received Signal Strength
(RSS) in the $\emph{Physical Interference Model}$ \cite{han2010automatic}\cite{graph-cluster}\cite{graph-dynamic},
and rules based on the locations in the $\emph{Protocol Model}$ \cite{jain2005impact}.
If one pair identifies another pair as strongly interfering, its decision
can be relayed by the managing servers to the latter, such that any
two pairs can reach consensus of whether there exists an edge between
them.

\textbf{Step 2. Distributed generation of MISs that span all the UEs:
}In Step 2, the UE-SBS pairs generate a subset of MISs in a distributed
fashion. It is important that the generated subset spans all the UEs,
namely every UE is contained in at least one MIS in the subset. Otherwise,
some UEs will never be scheduled.

The key idea is that from a given list of colors, each UE has to choose
a set of colors such that the choice does not conflict with its neighbors.
We should ensure that each UE has at least one color. We call the
set of UEs with the same color {}``a color class''. In addition,
we should also ensure that every color class is a MIS. This step is
composed of two phases: first, distributed coloring of the interference
graph based on \cite{johansson1999simple}, and second, extension
of color classes to MISs. All the UEs are synchronized and carry out
their computation simultaneously. We now explain the algorithm in
detail. The pseudo-codes can be found in Table II and III in the Appendix.
\begin{figure}
\centering{}\includegraphics[width=6in]{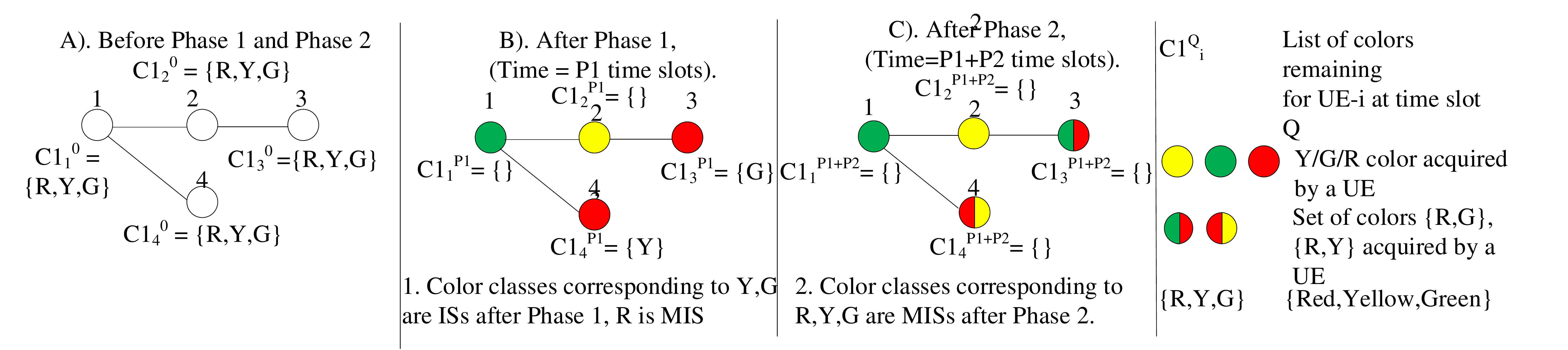}\caption{\label{fig:Illustration-of-Algorithm}Illustration of the distributed
generation of MISs in Step 2. }
\end{figure}

\textbf{Phase 1. Distributed coloring of the interference graph}:
Let $H$%
\footnote{The maximum number of colors $H$ should be set to be larger than
the maximum number of UE-SBS pairs interfering with any UE-SBS pair.
The SBSs can determine $H$ according to the deployment scenario.
$H$ in general will also include the number of UEs that use the same
SBS who interfere with each other along with the other neighboring
UEs. For example, $H$ can be 10-15 in an office building with dense
deployment of SBSs, and can be 3-5 in a residential area. %
} be the maximum number of colors given to all SBSs at the installation
and $d_{i}$ be the degree (number of neighbors in the interference
graph) of the $i^{th}$ pair. The goal of this phase is to let each
UE-SBS pair $i$ choose \textit{one} color from $C_{i}^{0}\triangleq\{1,...H\}\cap\{1,..,d_{i}+1\}$,
such that no neighbors choose the same color. The distributed coloring
works as follows.

i) At the beginning of each time slot $t$, each UE $i$ chooses a
color from the set of remaining colors $C_{i}^{t}$ uniformly randomly,
and informs its neighbors of its tentative choice. This information
can be transmitted through the back-haul network/X2 interface that
is used for ICIC \cite{lopez2011enhanced}.

ii) If the tentative choice of a UE does not conflict with any of
its neighbor, then it fixes its color choice and informs the neighbors
of its choice. This UE does not contend for colors any further in
Phase 1. The neighbors delete the color chosen by $i$ from their
lists $C_{_{j}}^{t+1},\forall j\in\mathcal{N}(i)$, where $\mathcal{N}(i)$
is the set of $i$'s neighbors.

iii) Otherwise, if there is a conflict, then the UE does not choose
that color and repeats i) and ii) in the next time slot. 

There are $\lceil c_{1}\log_{\frac{4}{3}}N\rceil+1$ time slots in
Phase 1, where $c_{1}$ is the parameter given by the protocol. The
number of time slots is known to the SBSs at installation. Phase 1
is successful if all the UEs acquire a color, which implies that the
set of color classes (i.e., the set of UE-SBS pairs with the same
color) spans all the UEs. 

%
{}

\textbf{Phase 2. Extending color classes to the MISs:} Each color
class obtained at the end of Phase 1 is an independent set (IS) of
the graph. In Phase 2, we extend each of these ISs to MISs and possibly
generate additional MISs. After Phase 1, each UE has chosen one color
and deleted some colors from its list. But there may still be remaining
colors in its list that are not acquired by any of its neighbors.
If the UEs can acquire these remaining colors without conflicting
with its neighbors, then each color class will be a MIS. Phase 2 works
as follows.

i) At each time slot in Phase 2, UE $i$ chooses each color from the
remaining colors in its list independently with probability $c$.
Each UE $i$ then sends the set of its tentative choices to its neighboring
UEs, and receives their neighbors' choices.

ii) For any tentative choice of color, if there is a conflict with
at least one neighbor, then that color is not fixed; otherwise, it
is fixed. 

iii) At the end of each time slot, each UE deletes its set of fixed
colors from its list, and transmits this set of fixed colors to its
neighbors, who will delete these fixed colors from their lists as
well. Note that a UE deletes a particular color if and only if the
UE itself or some of its neighbors have chosen this color. Based on
this key observation, we can see that if a color is not in any UE's
list, the set of UEs with this color is a MIS. If all the UEs have
an empty list, then for any color in the set $\{1,...,H\}$, the set
of UEs with this color is a MIS.

There are $\lceil c_{2}\log_{x}N\rceil+1$ time slots in Phase 2,
where $x=\frac{1}{1-(c)^{H}(1-c)^{H^{2}}}$, and $c_{2}$ is the
parameter given by the protocol. The number of time slots is known
to the SBSs at installation. We say that Phase 2 is successful, if
it finds $H$ MISs, or equivalently if all the UEs have an empty list.

\textbf{Example:} We illustrate Step 2 in a network of 4 UE-SBS pairs,
whose interference graph is shown in Fig. \ref{fig:Illustration-of-Algorithm}.
At the start, each UE-SBS pair has a list of 3 colors $\{\text{Red, Yellow, Green}\}$.
Phase 1 is run for $P1=\lceil c_{1}\log_{\frac{4}{3}}5\rceil$ time
slots. At the end of Phase 1, UE 1 and UE 2 acquire Green and $\text{Yellow}$
respectively, while UEs 3-4 acquire Red. Hence, UE 1 (UE 2) has an
empty list, as Green (Yellow) is acquired by itself and Red, Yellow
(Green) by its neighbors. UE 3 (UE 4) has Green (Yellow) color in
its list of remaining colors.%
{} At the end of Phase 1, the Red color class is a MIS, while the Yellow
and Green color classes are not. Phase 2 is run for $P2=\lceil c_{2}\log_{x}5\rceil+1$
time slots. UE 3 (UE 4) acquires the remaining color Green (Yellow).
At the end of Phase 2, the Green and Yellow color classes become MISs
too.

The next theorem establishes the high success probability of Step
2.

\textbf{Theorem 1. }For any interference graph with the maximum degree
$\Delta\leq H-1$, the proposed algorithm in Table II and III outputs
a set of $H$ MISs that span all the UEs in $(\lceil c_{1}\log_{\frac{4}{3}}N\rceil+\lceil c_{2}\log_{x}N\rceil+2)$
time slots with  a probability no smaller than $(1-\frac{1}{N^{c_{1}-1}})(1-\frac{1}{N^{c_{2}-1}})$,
where $c_{1}$ and $c_{2}$ are design parameters that trade-off
the run time and the success probability.

See the Appendix for detailed proofs. 

Theorem 1 characterizes the performance of our proposed algorithm,
in terms of the run time of the algorithm and the lower bound of the
success probability. When the parameters $c_{1}$ and $c_{2}$ are
larger, the lower bound of the success probability increases at the
expense of a longer run time. When the maximum degree of the interference
graph is larger, we need to set a higher $H$, which results in a
longer run time. This is reasonable, because it is harder to find
coloring and MISs when the number of interfering neighbors is higher.
Finally, we can see that the lower bound of the successful probability
is very high even under smaller $c_{1}$ and $c_{2}$, especially
if the number of UEs is large. Note that the exact successful probability
should depend on the probability $c$ in Phase 2, while the lower
bound in Theorem 1 does not. Hence, our lower bound is robust to different
system parameters. Note also that the interference graph here is
a bounded-degree graph since the maximum degree is bounded by a given
constant, $H-1$. The algorithms in \cite{ramaswami1989distributed}\cite{ephremides1990scheduling}
(require ordering of the vertices, work sequentially and have a higher
complexity) can be used to output the MISs spanning all the UEs for
arbitrary graphs. However, we will show in Theorem 5, that the restriction
to bounded-degree graphs is a must to ensure that the minimum throughput
requirement of each UE is satisfied for any MIS based policy.

%
{}

%
{}

%
{}

 %
{} 

{}

\textbf{Step 3. Distributed computation of the optimal fractions of
time for each MIS:} Let the set of MISs generated in Step 2 be $\{I_{1}^{'},...,I_{H}^{'}\}$.
In Step 3, the UE-SBS pairs compute the fractions of time allocated
to each MIS in a distributed manner.

When an MIS is scheduled, the UEs in this MIS transmit at their maximum
power levels, and the other UEs do not transmit. Define $R_{i}^{k}$
as the $\textit{instantaneous}$ throughput obtained by UE $i$ in
the MIS $I_{k}^{'},$ which can be calculated as $\log_{2}(1+\frac{g_{iT(i)}p_{i}^{I_{k}^{'}}}{\sum_{r=1,r\not=i}^{N}g_{rT(i)}p_{r}^{I_{k}^{'}}+\sigma_{T(i)}^{2}})$,
where $p_{i}^{I_{k}^{'}}=p_{i}^{max}$ if $i\in I_{k}^{'}$ and $p_{i}^{I_{k}^{'}}=0$
otherwise. To determine $R_{i}^{k}$, the UE needs to know the total
interference it experiences when transmitting in $I_{k}^{'}$. This
can be measured by having an initial cycle of transmissions of UEs
in each MIS in the order of the indices of MISs/colors. 

From now on, we assume that the network performance criterion $W(\bm{y})$
is concave in $\textbf{y}$ and is separable, namely $W(y_{1},...y_{N})=\sum_{i=1}^{N}W_{i}(y_{i})$.
Examples of separable criteria include weighted sum throughput and
proportional fairness. Our framework can also deal with max-min fairness$\min_{i}R_{i}(\boldsymbol{\pi})$,
although it is not separable (see the discussion in the Appendix)
The problem of computing the optimal fractions of time for the MISs
is given as follows:\vspace{-1.95em}

\begin{eqnarray*}
\textbf{Coupled Problem\,(CP)} & \max_{\alpha} & \sum_{i=1}^{N}W_{i}\left(\sum_{k=1}^{H}\alpha^{k}R_{i}^{k}\right)\\
 & \text{subject to } & \sum_{k=1}^{H}\alpha^{k}R_{i}^{k}\geq R_{i}^{min},\,\forall i\in\{1,..,N\}\\
 &  & \sum_{k=1}^{H}\alpha^{k}=1,\alpha^{k}\geq0,\,\forall k\in\{1,..,H\}\end{eqnarray*}

Each UE $i$ knows only its own utility function $W_{i}$ and minimum
throughput requirement $R_{i}^{min}$. Hence, it cannot solve the
above problem by itself. We will first reformulate the above problem
into a decoupled problem and then show that the reformulated problem
can be solved in a distributed manner. Let each UE $i$ have a local
estimate $\beta_{i}^{k}$ of the fractions of time allocated to each
MIS $I_{k}^{'}$ (including those MISs that UE $i$ does not belong
to). We impose an additional constraint that all the UEs' local estimates
are the same. Note that this constraint will be satisfied by our solution,
and is not an assumption. Such a constraint is still global, because
any two UEs, even if they are not neighbors, need to have the same
local estimate. Hence, global message exchange among any pair of UEs
is still needed to solve this problem with local estimates and global
constraints%
\footnote{If the UEs could exchange messages globally, i.e. broadcast messages
to all the UEs in the network, and if the network performance criterion
is strictly concave, we could use standard dual decomposition with
augmented Lagrangian in \cite{bertsekas1989parallel} to derive a
distributed algorithm. However, in large networks, the UEs cannot
exchange messages globally with other UEs, and the network performance
criterion may not be strictly concave (e.g., the weighted sum throughput
is linear).%
}. To avoid global message exchange, we reformulate the CP into a decoupled
problem (DP) that involves only local coupling among the neighbors
and can be solved with local message exchange using Alternating Direction
Method of Multipliers (ADMM) \cite{wei20131}.

%
{}

Now we reformulate the CP into a decoupled problem (DP) that involves
only local coupling among the neighbors and that can be solved by
Alternating Direction Method of Multipliers (ADMM) \cite{wei20131}.
If UE $i$ and $l$ are connected by an edge $(i,l)$ then for each
set $I_{k}^{'}$ define $\theta_{(i,l)i}^{k}=\beta_{i}^{k}$ and $\theta_{(i,l)l}^{k}=-\beta_{l}^{k}$,
note that these auxiliary variables are introduced to formulate the
problem into the ADMM framework \cite{wei20131}. Define a polyhedron
for each $i$, $\mathcal{T}_{i}=\{\bm{\beta_{i}}|\bm{\text{s.t. }1^{t}\beta_{i}=1},\bm{\beta_{i}\geq0,\,}\bm{R_{i}^{'}\beta_{i}\geq}R_{i}^{min}\},$
here $\bm{\beta_{i}}=(\beta_{i}^{1},...,\beta_{i}^{H})$ and $\bm{R_{i}}=(R_{i}^{1},...,R_{i}^{H})$
and $()^{'}$corresponds to the transpose. Let $\bm{\beta=(\beta_{1},...,\beta_{N})}\in\bm{\mathcal{T}}$,
where $\bm{\mathcal{T}=}\prod_{i=1}^{N}\mathcal{T}_{i}$ and $\prod$
corresponds to the Cartesian product of the sets. Also, let $\bm{\beta^{k}}=(\beta_{1}^{k},...,\beta_{N}^{k}),\,\forall k\in\{1,..,H\}$.
Define another polyhedron $\Theta_{(i,l)}^{k}=\{(\theta_{(i,l)i}^{k},\theta_{(i,l)l}^{k}):\,\theta_{(i,l)i}^{k}+\theta_{(i,l)l}^{k}=0,\,-1\leq\theta_{(i,l)s}^{k}\leq1,\forall s\in\{i,l\}\}$,
$\bm{\Theta^{k}}=\prod_{(i,l)\in E}\Theta_{(i,l)}^{k}$ here $E=(e_{1},..e_{M})$
is the set of all the $M$ edges in the interference graph. A vector
$\bm{\theta^{k}\in}\bm{\Theta^{k}}$ is written as $\bm{\theta^{k}=}(\theta_{e_{1},z(e_{1})}^{k},\theta_{e_{1},t(e_{1})}^{k},..,\theta_{e_{M},z(e_{M})}^{k},\theta_{e_{M},t(e_{M})}^{k})$,
here $z(e_{i}),\, t(e_{i})$ correspond to the  vertices in the edge,
$e_{i}$. Similarly define, $\bm{\theta=(\theta^{1},...,}\bm{\theta^{H})}\in\bm{\Theta}$
, where $\bm{\Theta=}\prod_{k=1}^{H}\bm{\Theta^{k}}$.\vspace{-1.95em}

\begin{eqnarray*}
\textbf{Decoupled Problem\,(DP)} & \min_{\bm{\beta\in\mathcal{T}},\bm{\theta\in\Theta}}-\sum_{i=1}^{N}\bm{W}\bm{_{i}(\bm{R_{i}}{}^{'}\beta}_{i})\\
 & \text{subject to }\bm{D^{k}}\bm{\beta^{k}-\theta^{k}=0},\,\forall k\in\{1,..,H\}\end{eqnarray*}

Here, $\bm{D^{k}}\in\mathbb{R}^{2M\times N}$, is a matrix in which
each row has exactly one non-zero element which is $1$ or $-1$.
 Each element of the matrix, $D_{vj}^{k}$ is evaluated as follows,
the index $v$ can be uniquely expressed in terms of quotient $q$
and the remainder $w$ as $v=2q+w$, and if $j\not=z(e_{q+1}),\, j\not=t(e_{q+1})$
then $D_{vj}^{k}=0$. If $w=1,\, j=z(e_{q+1}),\,$ then $D_{vj}^{k}=1$
else if  $w=0,\, j=z(e_{q+1})\,$then $D_{vj}^{k}=0$. Also, if $w=0,\, j=t(e_{q+1}),\,$
then $D_{vj}^{k}=-1$ else if $w=1,\, j=t(e_{q+1})\,$ then $D_{vj}^{k}=0$.

\textbf{Theorem 2: }For any connected interference graph, the coupled
problem (CP) is equivalent to the decoupled problem (DP).

The above theorem shows that the original problem (CP), which requires
global information and global message exchange to solve, is transformed
into an equivalent problem (DP), which as we will show, can be solved
in a distributed manner with local message exchange 

We denote the optimal solution to the DP by ${\bf W}_{\text{distributed}}^{G}$.
We associate with each constraint $D_{eq}^{k}\beta_{q}^{k}=\theta_{eq}^{k}$
 a dual variable $\lambda_{eq}^{k}$. The augmented Lagrangian for
DP is $L_{y}\left(\{\bm{\beta}_{i}\}_{i},\{\theta_{eq}^{k}\}_{k,e,q},\{\lambda_{eq}^{k}\}_{k,e,q}\right)=-\sum_{i=1}^{N}W_{i}(\bm{\beta}_{i}^{T}\bm{R}_{i})+\sum_{k=1}^{H}\sum_{e\in E}\sum_{q\in e}\left[\lambda_{eq}^{k}\left(D_{eq}^{k}\beta_{q}^{k}-\theta_{eq}^{k}\right)+\frac{y}{2}\left(D_{eq}^{k}\beta_{q}^{k}-\theta_{eq}^{k}\right)^{2}\right]$.
In the ADMM procedure (see Table IV in the Appendix), each UE $i$
solves for its optimal local estimates $\bm{\beta}_{i}(t)$ that maximizes
the augmented Lagrangian given the previous dual variables $\lambda_{ei}^{k}(t-1)$
and auxiliary variables $\theta_{ei}^{k}(t-1)$ . Then it updates
its dual variable $\lambda_{ei}^{k}(t)$ and auxiliary variable $\theta_{ei}^{k}(t)$
based on its local estimate $\beta_{i}^{k}(t)$ and its neighbor $j$'s
local estimate $\beta_{j}^{k}(t)$. This iteration of updating local
estimates, dual variables, and auxiliary variables is repeated $P$
times. Next, it is shown that this procedure will indeed converge.

%
{}

%
{}

\textbf{Theorem 3: }If DP is feasible%
\footnote{DP is feasible, if the feasible region resulting from the constraints
in DP is non-empty.%
}, then the ADMM algorithm in Table IV converges to the optimal value
$\textbf{W}_{\text{distributed}}^{G}$ with a rate of convergence
$\mathcal{O}(\frac{1}{P})$. %
{}

%
{}

\textbf{Step 4. Determining the cycle length and transmission times:
}At the end of Step 3, all the UEs have a consensus about the optimal
fractions of time allocated to each MIS, namely $\bm{\beta}_{i}^{*}=\bm{\gamma}^{*}=(\gamma_{1}^{*},...,\gamma_{H}^{*}),\,\forall i\in\{1,..,N\}$.
The MISs transmit in the order of their indices (i.e., $\{1,..,H\}$)
in cycles. In each cycle of transmission, MIS $I_{k}^{'}$ transmits
for $\left\lceil \frac{\gamma_{k}^{*}}{\min_{i\in{1,...,N}}\gamma_{i}^{*}}\times10^{d}\right\rceil $
slots, where we multiply by $10^{d}$ such that the rounding error
is reduced or eliminated in case that $\frac{\gamma_{k}^{*}}{\min_{i\in{1,...,N}}\gamma_{i}^{*}}$
is not an integer. 

\vspace{-1em}

\begin{figure}
\centering \includegraphics[width=4in]{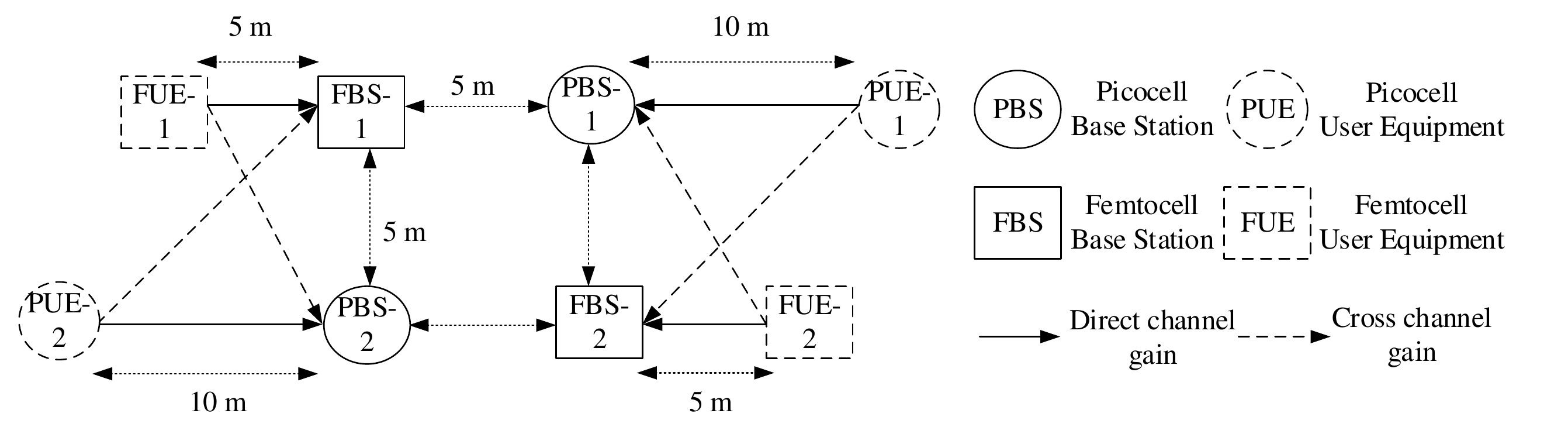}\caption{A heterogeneous network of 2 PBS and 2 FBS and their corresponding
UEs.}

\label{fig:1 user 4} %
\end{figure}

\subsection{A Motivation Example\label{sub:A-Motivating-Example}}

Consider a network of 2 picocell base stations (PBS) and 2 femtocell
base stations (FBS), each serving one UE. The network topology is
shown in Fig. 4. We assume a path loss model for channel gains, with
path loss exponent $4$. The maximum transmit power of each UE is
$80$ mW, and the noise power at each SBS is $1.6\times10^{-3}$ mW.
UEs in different tiers have different minimum throughput requirements:
FUE (femtocell UE) 1 and FUE 2 in the femtocells require a minimum
throughput $0.4$ bits/s/Hz, and PUE (picocell UE) 1 and PUE 2 in
the picocells require $0.2$ bits/s/Hz. The interference graph is
constructed according to a distance based threshold rule similar to
\cite{jain2005impact}. Specifically, an edge exists between two UE-BS
pairs if the distance between any pair of SBSs is less than a threshold,
which is set to be $1.2$m here. There are two MISs. MIS 1 consists
of FUE 1 and FUE 2, and MIS 2 consists of PUE 1 and PUE 2. We consider
two performance criteria: the max-min fairness and the sum throughput.
We will compare with the following state-or-the-art policies:

\textbf{1. Distributed Constant Power Control Policies}\cite{huang2006distributed,saraydar2001pricing,Decen-pow-game}\textbf{:
}In these policies, all the UEs choose $\emph{constant}$ power levels
determined by distributed algorithms utilizing  information (e.g.,
power levels used by neighbors) made available through local/global
message exchange. 

\textbf{2. Optimal Centralized Constant Power Policies: }In these
policies, all the UEs choose $\emph{constant}$ power levels determined
by a central controller utilizing global information. 

\textbf{3. Distributed MIS STDMA}-1\cite{ephremides1990scheduling}
and \textbf{STDMA}-2\cite{ramaswami1989distributed}\textbf{:} These
policies construct a subset of the MISs of the interference graph
in a distributed manner and propose fixed schedules of the MISs. Different
works adopt different schedules, and we differentiate them by referring
to them as MIS STDMA-1\cite{ephremides1990scheduling} and STDMA-2\cite{ramaswami1989distributed}.

\textbf{4. Distributed Joint Power Control and Spatial Reuse }\cite{kiani2008optimal}\cite{gesbert2007adaptation}\textbf{:
}These policies choose one UE from each cell to form a subset, and
schedule these subsets of UEs based on their channel gains to maximize
the sum throughput. The policies are named power matched scheduling
(PMS).  

In Table 1, we compare the performance of our proposed policy with
state-of-the-art policies for the same setup as in Fig. 4. We compute
the optimal centralized constant power control policy by exhaustive
search, which serves as the performance upper bound of the distributed
constant power control policies \cite{huang2006distributed,saraydar2001pricing,Decen-pow-game}
centralized constant power control policies \cite{chiang2007power}.
In PMS policies \cite{kiani2008optimal}\cite{gesbert2007adaptation},
UEs within the same cell are scheduled in a time-division multiple
access (TDMA) fashion, and the active UEs in different cells transmit
simultaneously. In this motivating example, there is one UE in each
cell, which will be scheduled to transmit all the time. Therefore,
the PMS policy reduces to a constant power control policy, and is
worse than the optimal centralized constant power control policy.
We can see that our proposed policy outperforms all constant power
control policies and distributed PMS policies by at least 375\% and
32.8\%, in terms of max-min fairness and sum throughput, respectively.
The significant performance improvement over the constant power control
policies results from the elimination of the high interference among
the users through scheduling MISs. Our proposed policy also outperforms
distributed STDMA policies by 30\%-40\%. As we will see in Section
VI, the performance gain is even higher (160\%-700\%) in realistic
deployment scenarios. Finally, in this motivating example, the proposed
policy achieves the optimal performance of the benchmark problem defined
in Section VI, which is a close approximation of the original problem
(CP). 

\begin{table}
\global\long\global\long\def\arraystretch{1.0}
 {\scriptsize {} \caption{Comparisons in terms of max-min fairness \& sum throughput criterion}
}{\scriptsize \par}

\centering{}{\scriptsize \label{table:5user_illun} \centering
}\begin{tabular}{|l|l|l|l|l|}
\hline 
{\scriptsize Policies}  & {\scriptsize Max-min}  & {\scriptsize Performance } & {\scriptsize Sum } & {\scriptsize Performance }\tabularnewline
 & {\scriptsize throughput (bits/s/Hz)} & {\scriptsize Gain \% } & {\scriptsize throughput (bits/s/Hz)}  & {\scriptsize Gain \% }\tabularnewline
\hline 
{\scriptsize Distributed constant power control }\cite{huang2006distributed,saraydar2001pricing,Decen-pow-game} & {\scriptsize <0.28} & {\scriptsize >375 \%} & {\scriptsize 6.1 } & {\scriptsize 32.8 \%}\tabularnewline
\hline 
{\scriptsize Distributed PMS }\cite{kiani2008optimal,gesbert2007adaptation}  & {\scriptsize <0.28 } & {\scriptsize >375\%} & {\scriptsize 6.1 } & {\scriptsize 32.8 \%}\tabularnewline
\hline 
{\scriptsize Optimal centralized constant power control}  & {\scriptsize 0.28} & {\scriptsize 375\%} & {\scriptsize 6.1}  & {\scriptsize 32.8 \%}\tabularnewline
\hline 
{\scriptsize Distributed MIS STDMA-2/1} \cite{ramaswami1989distributed,ephremides1990scheduling} & {\scriptsize 0.96}  & {\scriptsize 38.5\% } & {\scriptsize 6.25}  & {\scriptsize 30.0 \% }\tabularnewline
\hline 
{\scriptsize Proposed (Section-V}) & {\scriptsize 1.33}  & {-} & {\scriptsize 8.12}  & {-}\tabularnewline
\hline 
{\scriptsize Benchmark Problem (BP) (Section- VI)} & {\scriptsize 1.33}  & {-} & {\scriptsize 8.12}  & {-}\tabularnewline
\hline
\end{tabular}%
\end{table}

\vspace{-0.75em}

\subsection{Performance Guarantees for Large Networks and Properties of Interference
Graphs\label{sub:Performance-Guarantee-for}}

In this subsection, we provide performance guarantees for our proposed
framework described in Section \ref{sub:DesignFramework}. Specifically,
we prove that the network performance $\textbf{W}_{\text{distributed}}^{G}$
achieved by the proposed distributed algorithms has a constant competitive
ratio with respect to the optimal value $\textbf{W}_{\text{\text{opt}}}$
of the PDP. Moreover, we prove that the competitive ratio does not
depend on the network size. Our result is strong, because the solution
to PDP  needs to be computed by a centralized controller with global
information and with NP complexity, while our proposed framework allows
the UEs to compute the policy fast in a distributed manner with local
information and local message exchange.

Before characterizing the competitive ratio analytically, we define
some auxiliary variables. Define the upper and lower bounds on the
UEs' maximum transmit power levels and throughput requirements as,
$0<p_{lb}^{max}\leq p_{i}^{max}\leq p_{ub}^{max},\forall i\in\{1,...,N\}$
and, $0<R_{lb}^{min}\leq R_{i}^{min}\leq R_{ub}^{min},\forall i\in\{1,...,N\}$
respectively. Let $D_{ij}$ is the distance between UE $i$ and SBS
$j$. Define upper and lower bounds on the distance between any UE
and its serving SBS and the noise power at the SBSs as, $0<D^{lb}\leq D_{iT(i)}\leq D^{ub},\forall i\in\{1,...,N\}$
and, $\sigma_{lb}^{2}\leq\sigma_{j}^{2}\leq\sigma_{ub}^{2},\forall j\in\{1,...,K\}$
respectively. We assume that the channel gain is $g_{ij}=\frac{1}{(D_{ij})^{np}}$,
where $np$ is the path loss exponent. 

\textbf{Definition 1 (Weak Non-neighboring Interference):} The interference
graph $G$ exhibits  $\zeta$ Weak Non-neighboring Interference ($\zeta$-WNI)
if for each UE $i$ the maximum interference from its non-neighbors
is bounded, namely $\sum_{j\not\in\mathcal{N}(i),j\not=i}g_{jT(i)}p_{j}^{max}\leq(2^{\zeta}-1)\sigma_{ub}^{2},\,\forall i\in\{1,...,N\}$. 

Define $\Delta^{max}=\frac{\log_{2}(1+\frac{p_{lb}^{max}}{(D^{ub})^{np}2^{\zeta}\sigma_{ub}^{2}})}{R_{ub}^{min}}-1$.
Then we have the following theorem for the network performance criterion,
sum throughput%
\footnote{We can extend this result for weighted sum throughput, with weights
$w_{i}=\Theta(\frac{1}{N})$, it is not done to avoid complex notations.%
}. 

\textbf{Theorem 4:} For any connected interference graph, if  the
maximum degree $\Delta\leq\Delta^{max}$ and it exhibits $\zeta$-WNI
then, our proposed framework of interference management described
in Section \ref{sub:DesignFramework}   achieves a performance $\textbf{W}_{\text{distributed}}^{G}\geq\Gamma\cdot\textbf{W}_{\text{\text{opt}}}$
with a probability no smaller than $(1-\frac{1}{N^{c_{1}-1}})(1-\frac{1}{N^{c_{2}-1}})$.
Moreover, the competitive ratio $\Gamma=\frac{R_{ub}^{min}}{\log_{2}(1+\frac{p_{ub}^{max}}{(D^{lb})^{np}\sigma_{lb}^{2}})}$
is  independent of the network size.

Note that the analytical expression of competitive ratio, $\Gamma=\frac{R_{ub}^{min}}{\log_{2}(1+\frac{p_{ub}^{max}}{(D^{lb})^{np}\sigma_{lb}^{2}})}$,
does not depend on the size of the network. Our results are derived
under the conditions that the interference graph has a maximum degree
bounded by $\Delta^{max}$, and that the interference from non-neighbors
is bounded (i.e. $\zeta-$WNI). These conditions do not restrict the
size of the network, next example illustrates this. In addition, our
results hold for any interference graph that satisfy the conditions
in Theorem 4, regardless of how the graph is constructed.

$\emph{Example:}$ Consider a layout of SBSs in a $K\times K$ square
grid, i.e. $K^{2}$ SBSs with a distance of $5$m between the nearest
SBSs. Assume that each UE is located vertically below its SBS at a
distance of $1$ m. Fix the parameters $p_{i}^{max}=100$ mW, $\sigma_{i}^{2}=3$
mW, $R_{i}^{min}=0.1$bits/s/Hz, $\forall i\in\{1,..,K^{2}\}$, $np=4$.
We construct the interference graph based on the distance rule \cite{jain2005impact},
namely there is an edge between two pairs if the distance between
their SBSs exceeds 6m, which gives us the maximum degree $\Delta=4$.
We can also verify that the interference graphs under any number $K^{2}$
of SBSs exhibit $\zeta$-WNI with $\zeta=0.15$ and $\Delta<\Delta^{max}$,
where $\Delta^{max}=48$. Given $\Delta=4$ and $\zeta=0.15$, from
Theorem 4, we get the performance guarantee of $0.17$ for any network
size $K^{2}$. Note that the number $0.17$ is a performance guarantee,
and that the actual performance is much higher compared to the performance
guarantee as well as those achieved by state-of-the-art policies (see
Section \ref{sec:Illustrative-Results}).%
{}

%
{}

Both Theorem 1 and 4 required the maximum degree of the interference
graph to be bounded by a given constant. Here, we show that constraint
on the degree is natural and is a must to ensure feasibility, i.e.
to satisfy the minimum throughput requirements of every UE. Specifically,
we prove that if the maximum degree exceeds some threshold,  then
no policy based on scheduling MISs in $\Pi^{MIS}$ (a large space
of policies, see Section \ref{sec:System-Model-l}) is feasible. Let
the construction of interference graph be based on a distance based
threshold rule similar to \cite{jain2005impact}. An edge exists between
two UE-SBS pairs if and only if, the distance between two SBS is no
greater than $D^{th}$. We define the threshold of the maximum degree
as $\Delta^{*}$ (See the Appendix for the expression).%
{}

\textbf{Theorem 5: }If the maximum degree of the interference graph
$\Delta\geq\Delta^{*}$, then any policy based on scheduling MISs
in $\Pi^{MIS}$ fails to satisfy the minimum throughput requirements
of the UEs. 

The intuition behind Theorem 5 is that, if the degree of the interference
graph is large then there must be a large number of UE-SBS pairs which
interfere with each other strongly (mutually connected) which makes
it impossible to allocate each UE enough transmission time to satisfy
its minimum throughput requirement.

\vspace{-1em}

\subsection{Self-Adjusting Mechanism for Dynamic Entry/Exit of UEs\label{sub:Self-Adjusting-Mechanism-for}}

We now describe how the proposed framework can adjust to dynamic entry/exit
by the UEs in the network without recomputing all the four steps.
We allow the UEs to enter and exit, but number of SBSs is fixed. We
only allow let one UE enter or leave the network in any time slot. 

1. $\emph{UE leaves the network:}$ Suppose a UE $i$ which was transmitting
to SBS $T(i)$ leaves the network. If the UE $i$ was transmitting
in a set of colors $C{}_{i}$, then as soon as it leaves, these colors
can be potentially used by some neighbors, $\mathcal{N}(i)$. The
SBS $T(i)$ which was serving the UE $i$ can  have other UEs which
are still in the network and transmitting to it. Then for each color
$c'\in C{}_{i}$ it first searches among the UEs which it serves that
are not already transmitting in $c'$ and who also do not have a neighboring
UE-SBS pair which is already transmitting in $c'$. Let the set of
such UEs be $UE{}_{i,left}^{c'}$. SBS $T(i)$ allocates color $c'$
to the UE whose index is $\text{arg}\max_{j\in UE_{i,left}^{c'}}R_{j}^{c'}$.
In case $UE{}_{i,left}^{c'}$ is empty then that color, $c^{'}$ is
left unused.

2. $\emph{UE enters the network:}$ Suppose a UE $i$ registered with
SBS $T(i)$ enters the network. 

i). Given the minimum throughput requirement of the UE $i$ the SBS
$T(i)$ first creates a list of UEs, $UE_{i,enter}$, which consists
of the UEs it is serving and who are transmitting at more than their
minimum throughput requirement.

ii). SBS $T(i)$ creates the list of colors, $C{}_{i,enter}$ in which
UEs in $UE_{i,enter}$ are transmitting, it also consists of the colors
that are not being used by any UE served by $T(i)$. Next, it creates
valid colors list i.e.  $C{}_{i,enter}^{valid}$ from $C{}_{i,enter}$,
where a color $c\in C{}_{i,enter}^{valid}$  if $c\in C{}_{i,enter}$
and if none of the neighbors of $i$ in $\mathcal{N}(i)$ that are
not in $UE_{i,enter}$ are already using that color.

iii). Next, the SBS $T(i)$ has to allocate some portions from the
fractions of time allocated to the colors in $C{}_{i,enter}^{valid}$,
such that UE-$i$ can transmit and its minimum throughput requirement
is satisfied to the best possible extent. The allocation is done as
follows, let $C{}_{i,enter}^{valid}=\{c_{1}^{'},....,c_{s}^{'}\}$.
Proceeding sequentially, for each color $c_{i}^{'}$, SBS $T(i)$
selects the maximum possible portion to satisfy the minimum throughput
requirement of UE-$i$,  such that the minimum throughput requirements
of UEs in $UE_{i,enter}$, who are using this color, $c_{i}^{'}$
are not violated. 

iv). If the requirement of UE-$i$ is not satisfied then, SBS $T(i)$
requests the neighboring UE-SBSs (apart from the UEs that are served
by $T(i)$) to announce the set of colors which are either not being
used or in which their corresponding UEs are operating at more than
the minimum throughput requirement. From the set of colors that are
received, the SBS-$T(i)$ chooses each color from the list if it is
not being used by any other neighboring UE apart from the ones who
sent the announcement. The resulting list of colors is $Cl_{i,enter}^{valid}=\{c_{1}^{'},...,c_{l}^{'}\}$.

v). Proceeding sequentially with the colors in $Cl_{i,enter}^{valid}$,
for each color, SBS-$T(i)$ requests a portion from the fraction  of
time allocated to that color, to the neighboring UE-SBSs allocated
that color, such that the throughput requirement of UE-$i$ is satisfied.
The neighboring UE-SBSs either allow the requested portion or send
the portion which is acceptable to them, i.e. their throughput requirements
are not violated. SBS-$T(i)$ allocates the minimum acceptable portion
to UE-$i$ and proceeds to the next color in the list if the throughput
requirements are not satisfied. %
{}

%
{}%
{}

%
{}

\subsection{Extensions}

In our model, UEs operate in the same frequency band. However, our
methodology can be extended to scenarios where UEs operate in different
frequency channels (frequency reuse) and transmit at the same time.
In this case, the problem is to find the optimal frequency allocation
with the same objective function and constraints as in PDP. To solve
this problem, the first two steps of the framework remain the same.
In Step 3, the UEs compute distributedly the optimal fractions of
\textit{bandwidth} to be allocated to each MIS. This step is equivalent
to computing the optimal fraction of time allocated to each MIS as
in our current formulation. In Step 4, the UEs compute the number
of frequency channels allocated to each MIS based on the bandwidth
allocation. 

Note that we do not implement beamforming, although beamforming can
be used in conjunction with our policy. If the UEs transmitting to
the same SBS cooperate to do beamforming, we can delete the edge between
them in the interference graph, and use the new interference graph
in the scenario with beamforming. 

\vspace{-0.75em}

\section{Illustrative Results\label{sec:Illustrative-Results}}

In this section, we evaluate our proposed policy under a variety of
scenarios with different levels of interference, large numbers of
UEs, different performance criteria, time-varying channel conditions,
and dynamic entry and exit of UEs.

We compare our policy with the optimal centralized constant power
control policy, the distributed MIS STDMA-1 \cite{ephremides1990scheduling}
and STDMA-2 \cite{ramaswami1989distributed}, distributed PMS\cite{kiani2008optimal}\cite{gesbert2007adaptation},
in terms of sum throughput and max-min fairness. We do not separately
compare with distributed/centralized constant power control policies
in \cite{huang2006distributed,saraydar2001pricing,Decen-pow-game}\cite{chiang2007power},
because their performance is upper bounded by the optimal centralized
power control. Since it is difficult to compute the solution to
the NP-hard PDP, we  define a benchmark problem, where we restrict
our search to policies in which a UE either transmits at its maximum
power level or does not transmit.The space of such policies can be
writtenas $\Pi_{BC}=\left\{ \bm{\pi=}(\pi_{1},...,\pi_{N}):\pi_{i}:\mathbb{Z}_{+}\rightarrow\{0,p_{i}^{max}\}\,\forall i\in\{1,..,N\}\right\} $.
The policy space $\Pi_{BC}$ is a subset of all policies $\Pi$ and
is a superset of MIS based policies $\Pi^{MIS}$. In other words,
the benchmark problem has the same objective and constraints as PDP;
the only difference is the policy space to search . Hence, the benchmark
problem is a close approximation of the PDP. Note that the benchmark
problem is also NP-hard (see the appendix).

\vspace{-1.65em}

\subsection{Performance under time-varying channel conditions}

Consider a $3\text{x}3$ square grid of 9 SBSs with the minimum distance
between any two SBSs being $d=4.7$m. Each SBS serves one UE, who
has a maximum power of $1000$ mW and a minimum throughput requirement
of $0.45$ bits/s/Hz. The UEs and the SBSs are in two parallel horizontal
hyperplanes, and each SBS is vertically above its UE with a distance
of $\sqrt{10}m$. Then the distance from UE $i$ to another SBS $j$
is $D_{ij}=\sqrt{10+(D_{ij}^{BS})^{2}}$ , where $D_{ij}^{BS}$ is
the distance between SBSs $i$ and $j$. The channel gain from UE
$i$ to SBS $j$ is a product of path loss and Rayleigh fading  $f_{ij}\sim Rayleigh(\beta)$
, namely $g_{ij}=\frac{1}{(D_{ij})^{2}}f_{ij}$. The density function
of $Rayleigh(\beta)$ is $v(z)=\frac{z}{\beta^{2}}e^{-\frac{z^{2}}{2\beta^{2}}}$
for $\, z\geq0,$ and $v(z)=0$ for $z<0$. The SBSs identify neighbors
using a distance based rule with the threshold distance as in Section
V-C with $D^{th}=7$m. Note that different thresholds lead to different
interference graphs, and hence different performance, which will be
discussed next. Although, we use a distance based threshold rule,
our framework is general and does not rely on a particular rule. The
resulting interference graph for this setting is graph 3 shown in
Fig. 7 a)\@. 

\begin{figure}
\begin{centering}
\begin{minipage}[t]{0.45\textwidth}%
\includegraphics[width=3in,height=2.52in]{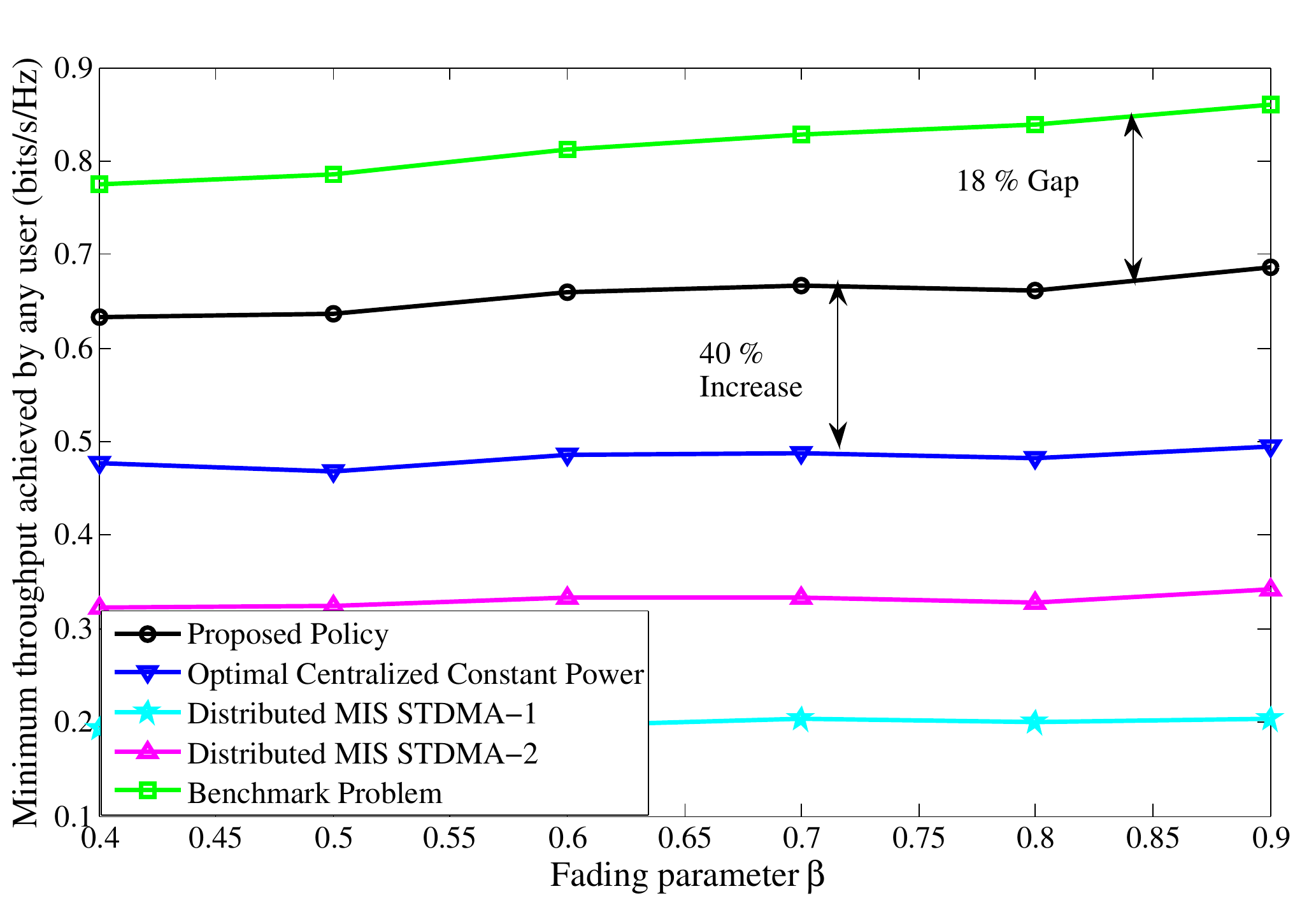}

\tiny\text{Fig. 5 a)}%
\end{minipage}\hfill{}%
\begin{minipage}[t]{0.45\textwidth}%
\includegraphics[width=3.4in,height=2.48in]{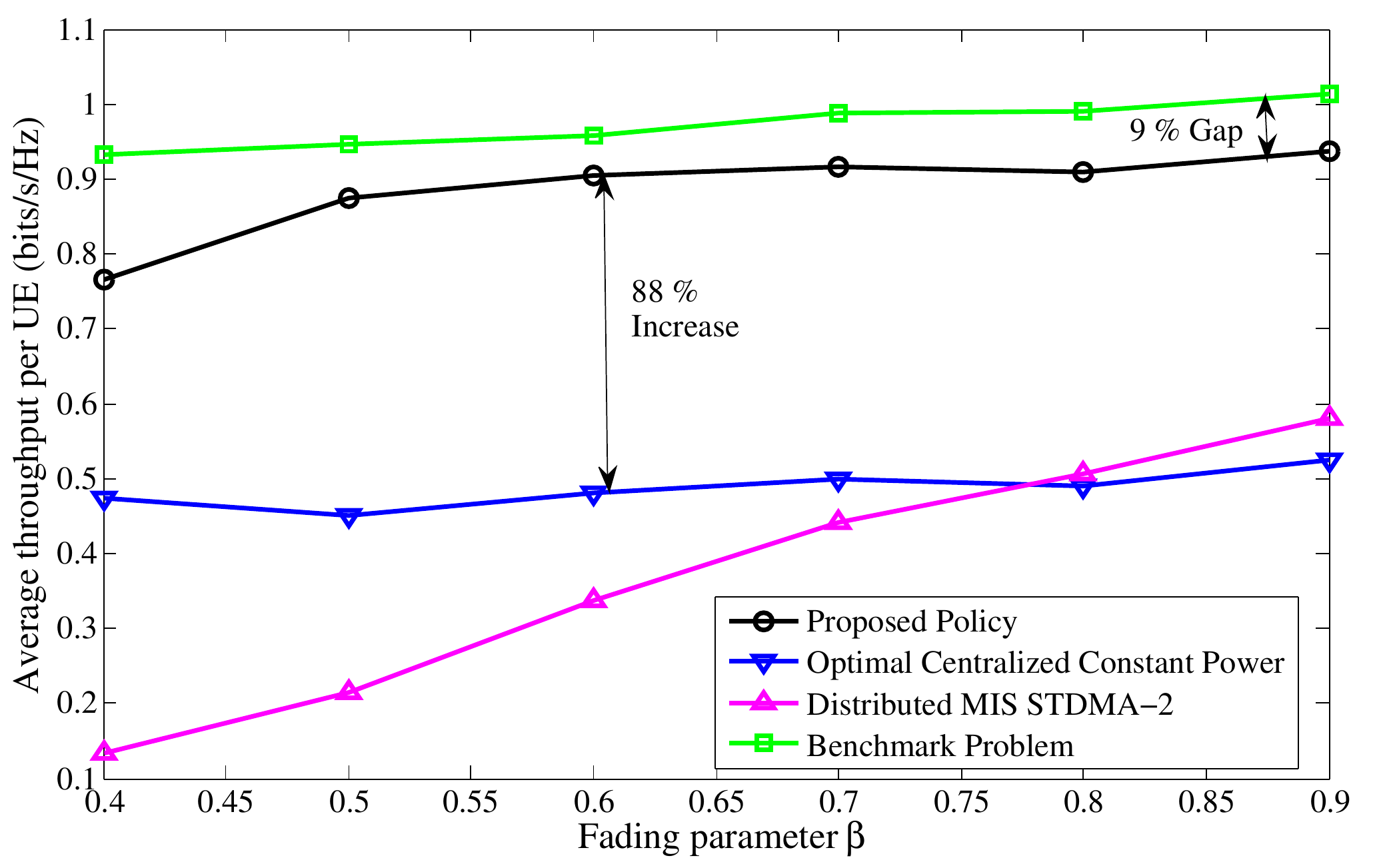}

\tiny\text{Fig. 5 b)}%
\end{minipage}\hfill{}
\par\end{centering}

\caption{\label{fig:Comparison-of-the}Comparison of the proposed policy with
state of the art under different interference strength and time-varying
channel conditions}
\end{figure}

At the beginning, the UE-SBS pairs generate the set of MISs (Step
2 of the design framework in Section V), and compute the optimal fractions
of time allocated to each MIS (Step 3). In our simulation, we assume
a block fading model \cite{goldsmith2005wireless} and the fading
changes every $100$ time slots independently. To reduce complexity,
the UEs do not recompute the interference graph and the MISs, but
will recompute the optimal fractions of time under the new channel
gains every 100 time slots. In Fig. \ref{fig:Comparison-of-the},
we compare the performance of the proposed policy with state of the
art policies under different variances $\beta$ of Rayleigh fading.
We do not plot the performance of distributed PMS for this scenario
since it is upper bounded by optimal centralized constant power control
(because there is one UE per cell). We do not plot the distributed
MIS STDMA -1 either, when the performance criterion is average throughput
per UE (i.e., $\frac{\text{sum throughput}}{N}$), because it cannot
satisfy the minimum throughput constraints. From Fig. \ref{fig:Comparison-of-the},
we can see that in terms of both average throughput and max-min fairness,
our proposed policy achieves large performance gain (up to 88\%) over
existing policies, and achieves performance close to the benchmark
(as close as 9\%).

\begin{figure}
\begin{centering}
\begin{minipage}[t]{0.45\textwidth}%
\includegraphics[width=3in]{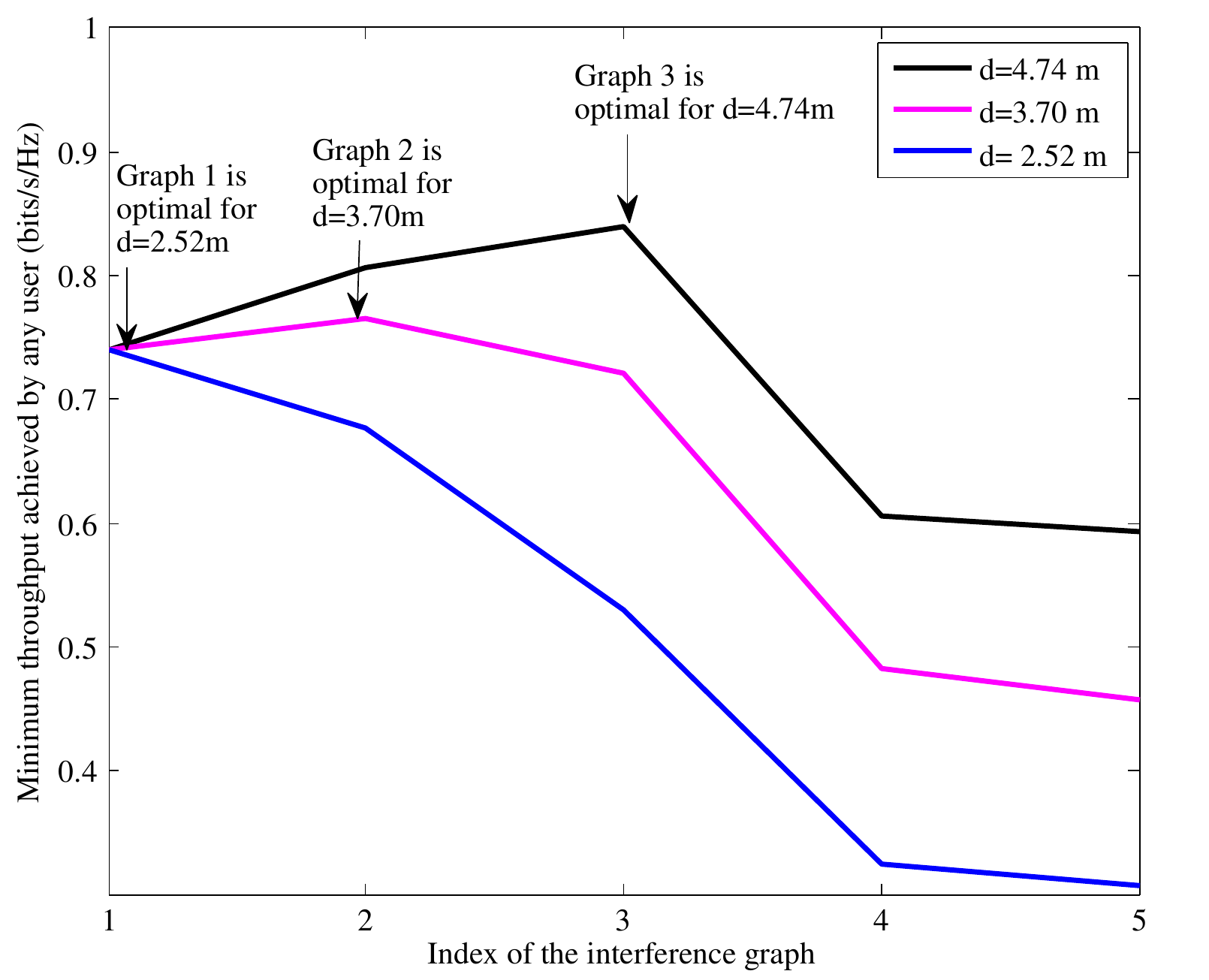}

\tiny\text{Fig. 6 a)}%
\end{minipage}\hfill{}%
\begin{minipage}[t]{0.45\textwidth}%
\includegraphics[width=3.4in,height=2.45in]{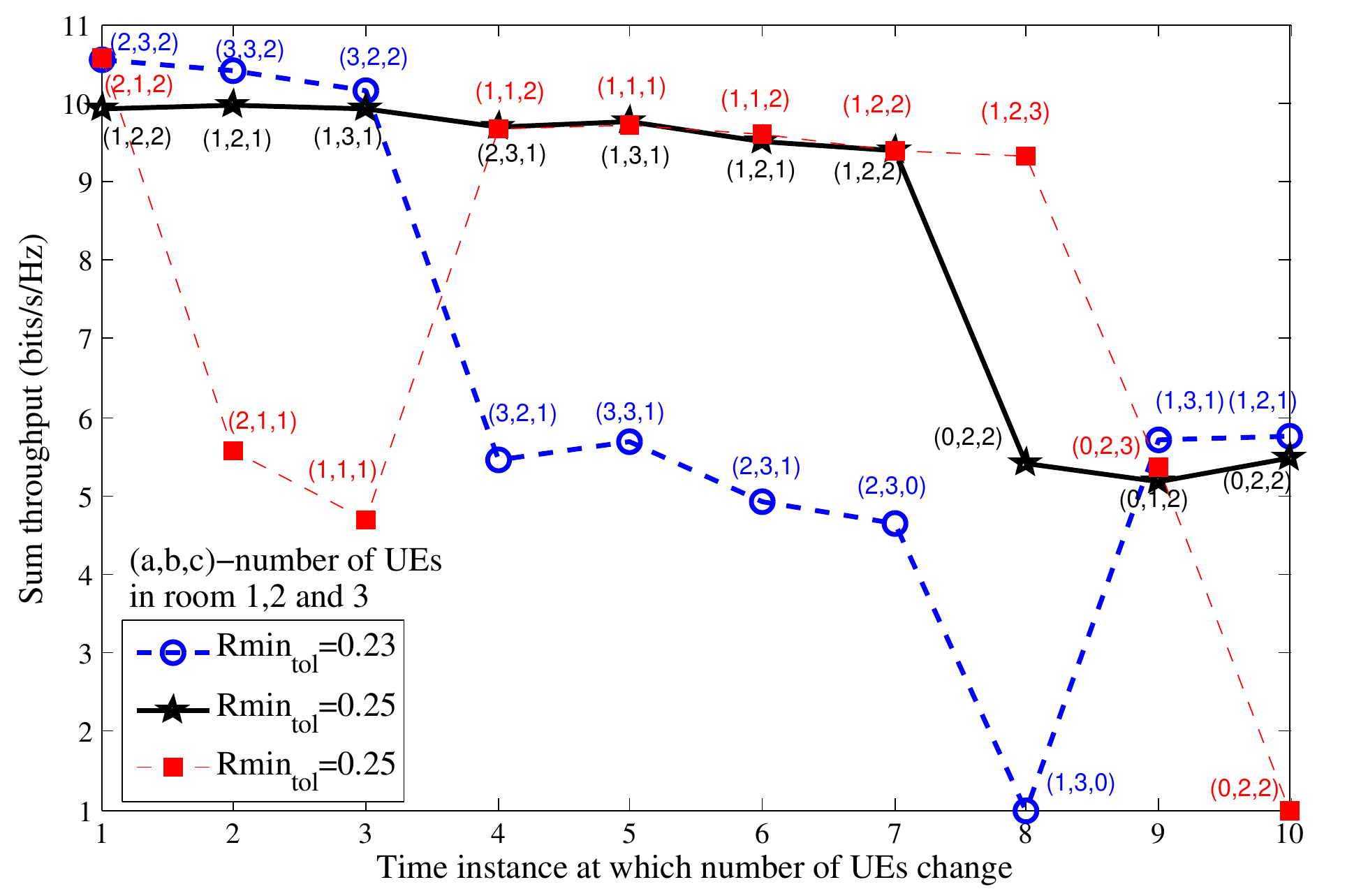}

\tiny\text{Fig. 6 b)}%
\end{minipage}\hfill{}
\par\end{centering}

\caption{a) Comparison of max-min fairness under different grid sizes, b) Sample
paths of sum throughput under dynamic entry/exit of UEs in the network}
\end{figure}

$\emph{Selecting the Optimal Interference Graph}:$ For different
values of $d$, there can be five possible interference graphs, which
are shown in Fig. 7 a). In Fig. 6 a) we show that as the grid size
$d$ decreases ($d=4.7m,d=3.7m$ and $d=2.5m$), the levels of interference
from the adjacent UEs increases, and as a result, the interference
graph with higher degrees perform better (as $d$ decreases, the optimal
graph changes from graph 3 to graph 1) .%
{}

\begin{figure}
\begin{minipage}[t]{0.55\textwidth}%
\includegraphics[width=3.3in,height=0.8in]{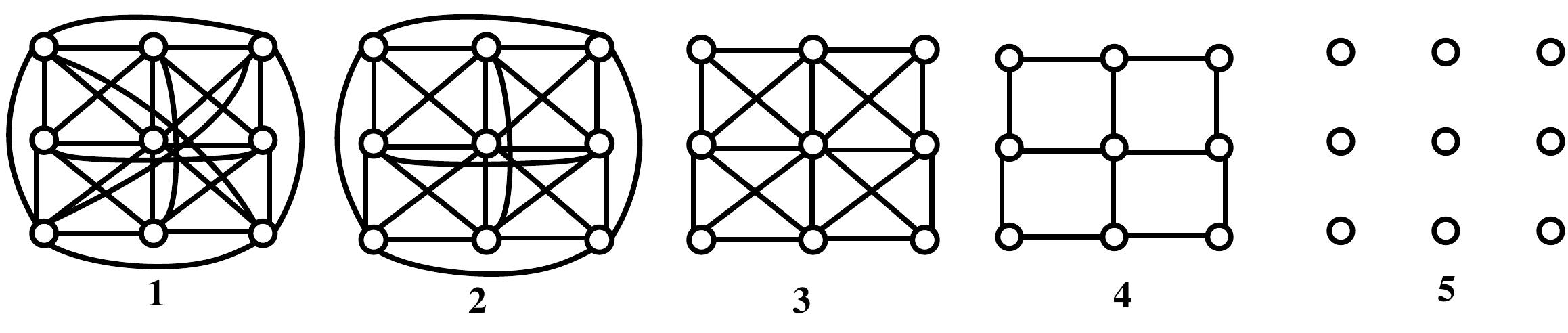}

\tiny\text{Fig. 7 a)}%
\end{minipage}%
\begin{minipage}[t]{0.35\textwidth}%
\includegraphics[width=1.7in]{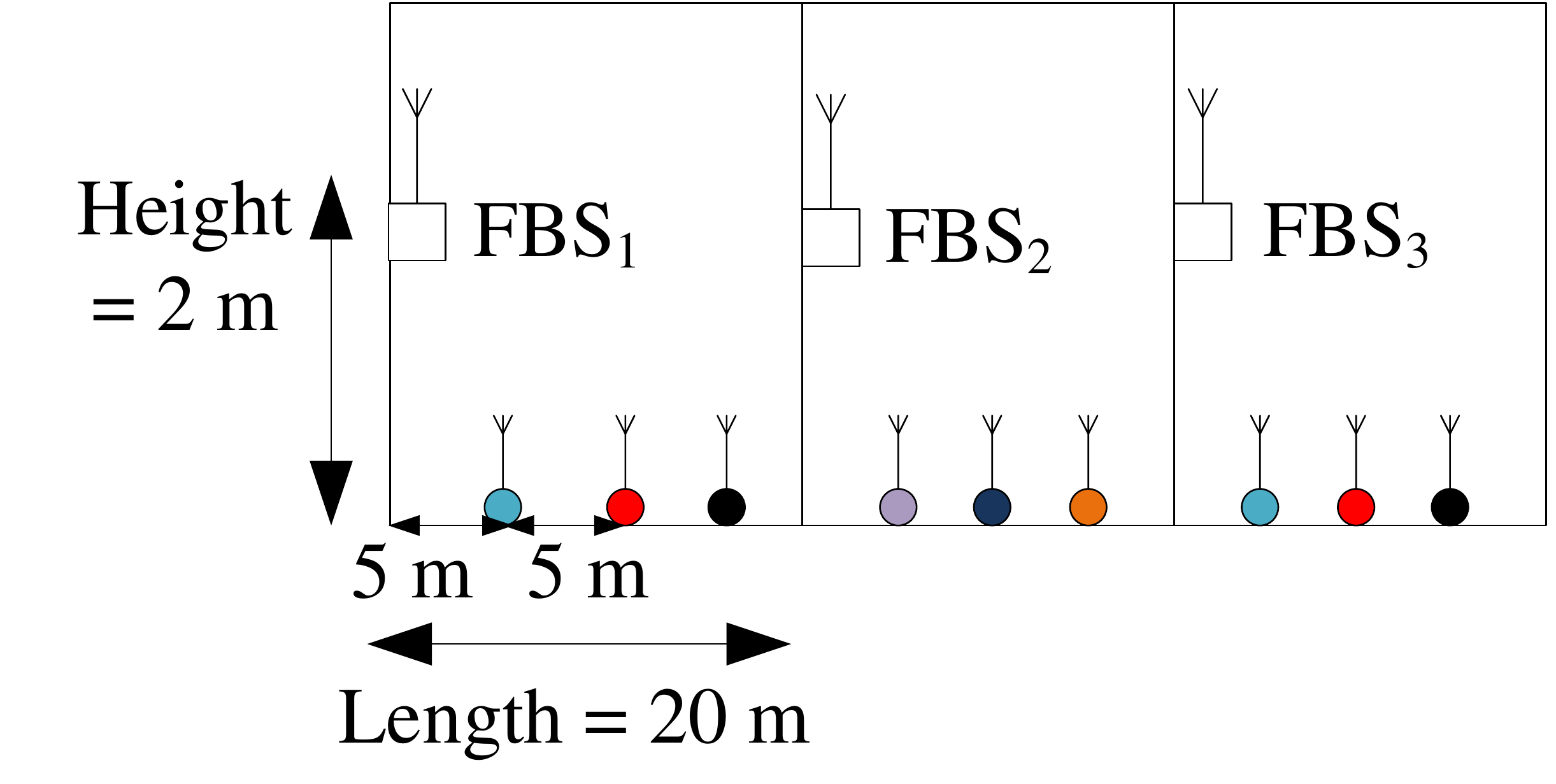}

\tiny\text{Fig. 7 b)}%
\end{minipage}\label{fig:int graph 9 user-1} \centering \caption{a) Different interference graphs for the 3 x 3 BS grid, b) Illustration
of setup with 3 rooms. }

%
\end{figure}

\vspace{-1.5em}

\subsection{Performance scaling in large networks\label{sub:Performance-scaling-in}}

\begin{figure}
\begin{centering}
\begin{minipage}[t]{0.45\textwidth}%
\includegraphics[width=3in]{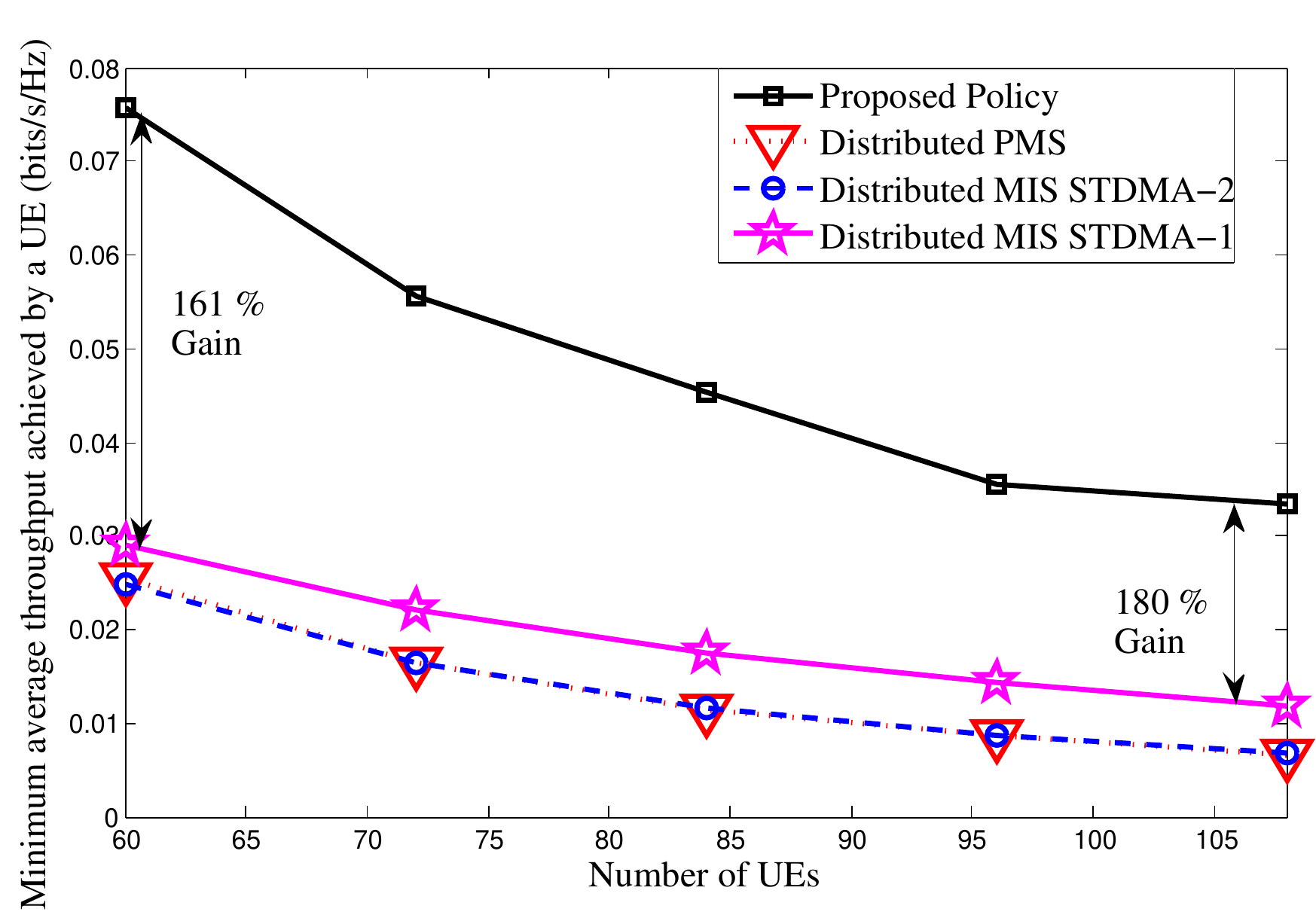}

\tiny\text{Fig. 8 a)}%
\end{minipage}\hfill{}%
\begin{minipage}[t]{0.45\textwidth}%
\includegraphics[width=3.4in,height=2in]{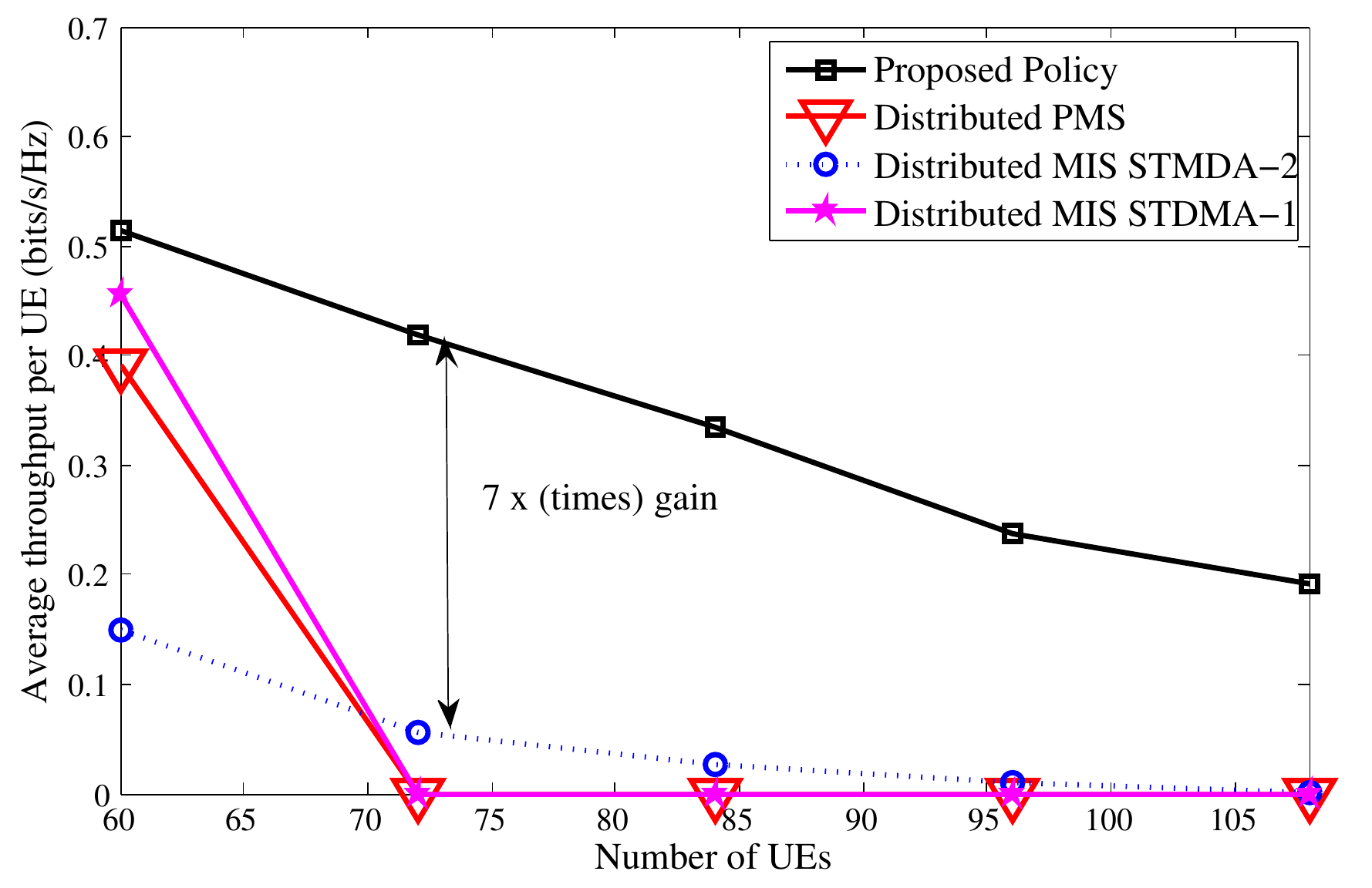}

\tiny\text{Fig. 8 b)}%
\end{minipage}\hfill{}
\par\end{centering}

\caption{\label{fig:Comparing-the-proposed}Comparison of max-min fairness
and average throughput per UE against state of the art for large networks}
\end{figure}

Consider the uplink of a femtocell network in a building with 12 rooms
adjacent to each other. Fig. 7 b) illustrates 3 of the 12 rooms with
5 UEs in each room. For simplicity, we consider a 2-dimensional geometry.
Each room has a length of 20 meters. In each room, there are $P$
uniformly spaced UEs, and one SBS installed on the left wall of the
room at a height of 2m. The distance from the left wall to the first
UE, as well as the distance between two adjacent UEs in a room, is
$\frac{20}{(1+P)}$ meters. Based on the path loss model in \cite{seidel1992914},
the channel gain from each SBS $i$ to a UE $j$ is $\frac{1}{(D_{ij})^{2}\Delta^{n_{ij}}}$,
where $\Delta=10^{0.25}$ is the coefficient representing the loss
from the wall, and $n_{ij}$ is the number of walls between UE $i$
and SBS $j$. Each UE has a maximum transmit power level of $50$
mW, a minimum throughput requirement of $R_{i}^{min}=0.025$ bits/s/Hz,
and a noise power level of $10^{-11}$mW at its receiver. Here, we
consider that the UEs use a distance based threshold rule as in Section
V-B with $D^{th}=30$ m. This results in interference graphs which
connects all the UE-SBS pairs within the room and in the adjacent
rooms. We vary the number $P$ of UEs in each room from $5$ to 9
 and compare the performance in Fig. \ref{fig:Comparing-the-proposed}.
Note that the optimal centralized constant power policy cannot satisfy
the feasibility conditions for any number of UEs in each room. Hence,
only the performance of distributed MIS STDMA-1,2 and distributed
PMS is shown in Fig. \ref{fig:Comparing-the-proposed}. We can see
that under both criteria, the performance gain of our proposed policy
is significant (from 160\% to 700\%). Note that since the number of
UEs is large, it is impossible to solve the benchmark problem (which
is NP-hard) is not possible. 

%
{}

\vspace{-1.25em}

\subsection{Self-adjusting mechanism for dynamic entry/exit of the UEs}

The self-adjusting mechanism proposed in Section \ref{sub:Self-Adjusting-Mechanism-for}
is aimed to provide incoming UEs with the maximum possible throughput
without affecting the incumbent UEs, and to reuse the time slots left
vacant by exiting UEs efficiently. Consider the same setup as in Section
\ref{sub:Performance-scaling-in} with 3 rooms and a maximum of $P=3$
UEs in each room. Each UE has a maximum transmit power of $1000$
mW and a minimum throughput requirement of $0.25$ bits/s/Hz.

%
{} We assume that at a given time only one UE either enters or leaves
the network%
{}.%
{} In Fig. 6 b) we show different sample paths of the sum throughput
under different entry and exit processes. In the legends (i.e., $Rmin_{tol}$),
we show the minimum throughput achieved at any point in the sample
path.  We repeated the same procedure 100 times. We can see that the
self-adjusting mechanism works well by guaranteeing a worst-case minimum
throughput requirement of $0.23$ bits/s/Hz, which is just $0.02$
bits/s/Hz below the original requirement more than $80$\% of the
time. 

%
{}

\vspace{-0.9em}

\section{Conclusion \label{sec:Conclusion}}

We proposed a design framework for distributed interference management
 in large-scale, heterogeneous networks, which are composed of different
types of cells (e.g. femtocell, picocell), have different number of
UEs in each cell, and have UEs with different minimum throughput requirements
and channel conditions. Our framework allows each UE to have only
local knowledge about the network and communicate only with its interfering
neighbors. There are two key steps in our framework. First, we propose
a novel distributed algorithm for the UEs to generate a set of MISs
that span all the UEs. The distributed algorithm for generating MISs
requires $\mathcal{O}(\log N)$ steps (which is much faster than state-of-the-art)
before it converges to the set of MISs with a high probability. Second,
we reformulate the problem of determining the optimal fractions of
time allocated to the MISs in a novel manner such that the optimal
solution can be determined by a distributed algorithm based on ADMM.
Importantly, we prove that under wide range of conditions, the proposed
policy can achieve a constant competitive ratio with respect to the
policy design problem which is NP-hard. Moreover, we show that our
framework can adjust to UEs entering or leaving the network. Our simulation
results show that the proposed policy can achieve large performance
gains (up to $85$\%).

%
{}


%
\begin{table}
\caption{Generating MISs in a distributed manner, algorithm for UE $i$}

\centering{}\begin{tabular}{|l|}
\hline 
\textbf{Phase 1}- \textbf{Initialization:} $\text{Tx}_{\text{tent}}^{i}=\phi$,
$\text{Tx}_{\text{final}}^{i}=\phi$, tentative and final choice of
UE $i$, $\text{Rx}_{\text{tent}}^{\mathcal{N}(i)}=\phi$ ,$\text{Rx}_{\text{final}}^{\mathcal{N}(i)}=\phi$
tentative \tabularnewline
and final choice made by the neighbors, $C_{i}^{0}=\{1,...,H\}\cap\{1,..,d_{i}+1\}$
the current list of subset of available colors, \tabularnewline
$C_{i}=\phi,\text{list of colors used by \ensuremath{i}, F}_{\text{colored}}^{i}=\phi$,
$C1_{i}^{0}=\{1,...,H\},$the current list of all available colors\tabularnewline
\hline 
for $n=0$ to $\lceil c_{1}\log_{\frac{4}{3}}N\rceil$\tabularnewline
$\text{Tx}_{\text{tent}}^{i}=\phi$, $\text{Tx}_{\text{final}}^{i}=\phi$\tabularnewline
if($\text{F}_{\text{colored}}^{i}=\phi$)\tabularnewline
$\text{\,\,\ Tx}_{\text{tent}}^{i}=\text{rand\{}C_{i}^{n}\}$, rand
represents randomly selecting a color and informing the neighbors
about it.\tabularnewline
~~$\text{Rx}_{\text{\text{tent}}}^{\mathcal{N}(i)}=\{\text{Tx}_{\text{tent}}^{k},\forall k\in\mathcal{N}(i)\}$\tabularnewline
~~~~~If($\text{Tx}_{\text{tent}}^{i}\not=\text{Rx}_{\text{tent}}^{\mathcal{N}(i)}(j)$,~$\forall j\,\in\mathcal{N}(i)$),
here UE-$i$ checks if there is a conflict with any of the neighbor's
choice\tabularnewline
~~~~~~~~$\text{Tx}_{\text{final}}^{i}=\text{Tx}_{\text{tent}}^{i}$,
$C_{i}=\{\text{Tx}_{\text{final}}^{i}\}$,if there is no conflict
then UE-$i$ transmits its final color choice to the neighbors, \tabularnewline
~~~~~else\tabularnewline
~~~~~~~~$\text{Tx}_{\text{final}}^{i}=\phi$\tabularnewline
~~~~end\tabularnewline
end\tabularnewline
$\text{Rx}_{\text{\text{final}}}^{\mathcal{N}(i)}=\{\text{Tx}_{\text{final}}^{k},\forall k\in\mathcal{N}(i)\}$\tabularnewline
$C_{i}^{n+1}=C_{i}^{n}\cap\text{\{Rx}_{\text{final}}^{\mathcal{N}(i)}\cup\text{\text{Tx}}_{\text{final}}^{i}\}^{c}$,
$C1_{i}^{n+1}=C1_{i}^{n}\cap\text{\{Rx}_{\text{final}}^{\mathcal{N}(i)}\cup\text{\text{Tx}}_{\text{final}}^{i}\}^{c}$ \tabularnewline
if($\text{Tx}_{\text{final}}^{i}\not=\phi$)\tabularnewline
$\text{F}_{\text{colored}}^{i}=1$\tabularnewline
end\tabularnewline
end\tabularnewline
\hline
\end{tabular}%
\end{table}
\begin{table}
\caption{Phase 2 of the distributed MIS generation}

\centering{}\begin{tabular}{|l|}
\hline 
\textbf{Phase 2-Initialization: }$\text{Tx}_{\text{tent},i}^{\text{set}}=\phi$,$\text{Tx}_{\text{final},i}^{\text{set}}=\phi$,
the set of tentative and final colors chosen by $i$,$\text{Rx}_{\text{tent},i}^{\text{set}}=\phi$,\tabularnewline
 $\text{Rx}_{\text{final},i}^{\text{set}}=\phi$, the set of tentative
and final colors chosen that are received from the neighbors, $x=\frac{1}{1-(c)^{H}(1-c)^{H^{2}}}$\tabularnewline
\hline
for $n=\lceil c_{1}\log_{\frac{4}{3}}N\rceil+1$ to $\lceil c_{1}\log_{\frac{4}{3}}N\rceil+\lceil c_{2}\log_{x}N\rceil+1$ \tabularnewline
$\text{Tx}_{\text{tent},i}^{\text{set}}=\phi$,$\text{Tx}_{\text{final},i}^{\text{set}}=\phi$,\tabularnewline
~~~for $m=1$ to $|C1_{i}^{n}|$ \tabularnewline
~~~with probability $c$, $\text{Tx}_{\text{tent,i}}^{\text{set}}(m)=C1_{i}^{n}(m)$,
randomly selecting and informing the neighbors about tentative choice\tabularnewline
~~~with probability $1-c$, $\text{Tx}_{\text{tent,i}}^{\text{set}}(m)=\phi$ \tabularnewline
~~~end\tabularnewline
$\text{Rx}_{\text{tent},i}^{\text{set}}=\cup_{k\in\mathcal{N}(i)}\text{Tx}_{\text{tent,k}}^{\text{set}},$
set of tentative color choices of the neighbors of $i$\tabularnewline
~~~for $r=1$ to $|\text{Tx}_{\text{tent,i}}^{\text{set}}|$\tabularnewline
~~~If($\text{Tx}_{\text{tent},i}^{\text{set}}(r)\not=\text{Rx}_{\text{tent},i}^{\text{set}}(j)\,\forall j\in\mathcal{N}(i)$~)\tabularnewline
~~~$\text{Tx}_{\text{final},i}^{\text{set}}(r)=\text{Tx}_{\text{tent},i}^{\text{set}}(r)$\tabularnewline
~~~else\tabularnewline
~~~$\text{Tx}_{\text{final},i}^{\text{set}}(r)=\phi$\tabularnewline
~~~end\tabularnewline
$C_{i}=C_{i}\cup\text{Tx}_{\text{final},i}^{\text{set}}$\tabularnewline
$\text{Rx}_{\text{final},i}^{\text{set}}=\cup_{k\in\mathcal{N}(i)}\text{Tx}_{\text{final,k}}^{\text{set}}$,
set of final color choices of the neighbors of $i$\tabularnewline
$C1_{i}^{n+1}=C1_{i}^{n}\cap\text{\{Rx}_{\text{final},i}^{\text{set}}\cup\text{Tx}_{\text{final,i}}^{\text{set}}\}^{c}$ \tabularnewline
end\tabularnewline
\hline
\end{tabular}%
\end{table}

\vspace{-1.15em}

\begin{table}
\caption{ADMM update algorithm for UE $i$}

\centering{}\begin{tabular}{|l|}
\hline 
Initialization: arbitrary $\bm{\beta}_{i}(0)\in\mathcal{B}_{i}$,
$\theta_{ei}^{k}(0)$ such that $\bm{\theta}^{k}\in\bm{\Theta}^{k}$,
and $\lambda_{ei}^{k}(0)=0$, $\forall k\in\{1,...,H\}$,$\forall e$
such that $i\in e$\tabularnewline
\hline 
For $t=0$ to $P-1$\tabularnewline
$\bm{\beta}_{i}(t+1)=\text{arg}\min_{\bm{\beta}_{i}\in\mathcal{B}_{i}}-\sum_{i=1}^{N}W_{i}(\bm{\beta}_{i}^{T}\bm{R}_{i})+\sum_{k=1}^{H}\sum_{e\in E}\sum_{q\in e}\left[\lambda_{eq}^{k}\left(D_{eq}^{k}\beta_{q}^{k}-\theta_{eq}^{k}\right)+\frac{y}{2}\left(D_{eq}^{k}\beta_{q}^{k}-\theta_{eq}^{k}\right)^{2}\right]$\tabularnewline
$\bm{\beta}_{i}(t+1)$ is transmitted to all of its neighbors in $\mathcal{N}(i)$. \tabularnewline
$\lambda_{ei}^{k}(t)$ is transmitted to its neighbor connected with
edge $e$, $\forall k\in\{1,...,H\}$ and $\forall e$ such that $i\in e$\tabularnewline
Update $\forall k\in\{1,...,H\}$ and $\forall e$ such that $i\in e$\tabularnewline
$\lambda_{ei}^{k}(t+1)=\frac{1}{2}(\lambda_{ei}^{k}(t)+\lambda_{ej}^{k}(t))-\frac{y}{2}(D_{ei}^{k}\beta_{i}^{k}(t+1)+D_{ej}^{k}\beta_{j}^{k}(t+1))$,
where $j$ is the other endpoint of $e$.\tabularnewline
$\theta_{ei}^{k}(t+1)=\frac{1}{y}(\lambda_{ei}^{k}(t+1)-\lambda_{e,i}^{k}(t))+D_{ei}^{k}\beta_{i}^{k}(t+1)$\tabularnewline
end\tabularnewline
\hline
\end{tabular}%
\end{table}

\section*{Appendix}

\textbf{Discussion on max-min fairness: }We now discuss as to how
the proposed framework can be extended to incorporate inseparable
function like max-min fairness. The coupled problem with max-min fairness
objective is restated below: 

\begin{eqnarray*}
\textbf{Coupled Problem\,(CP)} & \max_{\bm{\alpha}} & \min_{i\in\{1,..,N\}}W_{i}(\sum_{k=1}^{H}\alpha_{k}R_{i}^{k})\\
 & \text{subject to} & \sum_{k=1}^{H}\alpha_{k}R_{i}^{k}\geq R_{i}^{min},\,\forall i\in\{1,...N\}\\
 &  & \sum_{k=1}^{H}\alpha_{k}=1,\,\alpha_{k}\geq0,\;\forall k\in\{1,...,H\}\end{eqnarray*}

Transforming the above problem into an equivalent problem with auxiliary
variable $t$ is given as 

\begin{eqnarray*}
\textbf{} & \max_{\bm{\alpha,}t} & t\\
 & \text{subject to} & W_{i}(\sum_{k=1}^{H}\alpha_{k}R_{i}^{k})\geq t,\,\forall i\in\{1,...,N\}\\
 &  & \sum_{k=1}^{H}\alpha_{k}R_{i}^{k}\geq R_{i}^{min},\,\forall i\in\{1,...N\}\\
 &  & \sum_{k=1}^{H}\alpha_{k}=1,\,\alpha_{k}\geq0,\;\forall k\in\{1,...,H\}\end{eqnarray*}

To decouple the above problem, we introduce local variables for each
UE $i$ given as,$\{\beta_{i}^{1},...,\beta_{i}^{H+1}\}$. Now we
state a problem which we claim is equivalent to CP,(the proof to this
claim is very similar to the proof of Theorem 2 and we will highlight
this fact in the proof clearly). 

\begin{eqnarray*}
\textbf{\textbf{P1}} & \max_{\bm{\beta}} & \sum_{i=1}^{N}\beta_{i}^{H+1}\\
 & \text{subject to} & W_{i}(\sum_{k=1}^{H}\beta_{i}^{k}R_{i}^{k})\geq\beta_{i}^{H+1},\,\forall i\in\{1,...,N\}\\
 &  & \sum_{k=1}^{H}\beta_{i}^{k}R_{i}^{k}\geq R_{i}^{min},\,\forall i\in\{1,...N\}\\
 &  & \sum_{k=1}^{H}\beta_{i}^{k}=1,\,\beta_{i}^{k}\geq0,\;\forall k\in\{1,...,H\},\forall i\in\{1,...,N\}\\
 &  & \beta_{i}^{k}=\beta_{j}^{k},\forall j\in\mathcal{N}(i),\forall k\in\{1,...,H+1\}\end{eqnarray*}

Here, $\bm{\beta}=(\bm{\beta_{1},..,\beta_{N}})$, with $\bm{\beta}_{i}=(\beta_{i}^{1},...,\beta_{i}^{H+1}),\forall i\in\{1,...,N\}$.
Now, given the two problems CP and the problem P1 are equivalent,
we focus on solving P1. P1 can be changed to a problem similar to
DP. To do that we introduce some additional variables similar to the
ones introduced for DP. If UE $i$ and $l$ are connected by an edge
$(i,l)$ then for each set $I_{k}^{'}$ define $\theta_{(i,l)i}^{k}=\beta_{i}^{k}$
and $\theta_{(i,l)l}^{k}=-\beta_{l}^{k}$, note that these auxiliary
variables are introduced to formulate the problem into the ADMM framework
\cite{wei20131}. Define a polyhedron for each $i$, $\mathcal{T}_{i}^{'}=\{\bm{(\beta1)_{i}}|\bm{\text{s.t. }1^{t}(\beta{}_{i}^{''})=1},\bm{(\beta1)_{i}\geq0,\,}\bm{R_{i}^{'}(\beta_{i}^{''})\geq}R_{i}^{min},W_{i}(\bm{R_{i}^{'}(\beta_{i}^{''})})-\beta_{i}^{H+1}\geq0\},$
here $\bm{\beta_{i}^{''}}=(\beta_{i}^{1},...,\beta_{i}^{H})$ and
$\bm{R_{i}}=(R_{i}^{1},...,R_{i}^{H})$ and $()^{'}$corresponds to
the transpose. Let $\bm{\beta=(\beta_{1},...,\beta_{N})}\in\bm{\mathcal{T}}^{'}$,
where $\bm{\mathcal{T}^{'}=}\prod_{i=1}^{N}\mathcal{T}_{i}^{'}$ and
$\prod$ corresponds to the Cartesian product of the sets. Also, let
$\bm{\beta^{k}}=(\beta_{1}^{k},...,\beta_{N}^{k}),\,\forall k\in\{1,..,H\}$.
Define another polyhedron $\Theta_{(i,l)}^{k}=\{(\theta_{(i,l)i}^{k},\theta_{(i,l)l}^{k}):\,\theta_{(i,l)i}^{k}+\theta_{(i,l)l}^{k}=0,\,-1\leq\theta_{(i,l)s}^{k}\leq1,\forall s\in\{i,l\}\}$,
$\bm{\Theta^{k}}=\prod_{(i,l)\in E}\Theta_{(i,l)}^{k}$ here $E=(e_{1},..e_{M})$
is the set of all the $M$ edges in the interference graph. A vector
$\bm{\theta^{k}\in}\bm{\Theta^{k}}$ is written as $\bm{\theta^{k}=}(\theta_{e_{1},z(e_{1})}^{k},\theta_{e_{1},t(e_{1})}^{k},..,\theta_{e_{M},z(e_{M})}^{k},\theta_{e_{M},t(e_{M})}^{k})$,
here $z(e_{i}),\, t(e_{i})$ correspond to the  vertices in the edge,
$e_{i}$. Similarly define, $\bm{\theta=(\theta^{1},...,}\bm{\theta^{H+1})}\in\bm{\Theta^{'}}$
, where $\bm{\Theta^{'}=}\prod_{k=1}^{H+1}\bm{\Theta^{k}}$. The reformulated
problem is stated as follows: 

\begin{eqnarray*}
\textbf{\,\ DP1} & \min_{\bm{\beta\in\mathcal{T}}^{'},\bm{\theta\in\Theta^{'}}}-\sum_{i=1}^{N}\bm{W}\bm{_{i}(\bm{R_{i}}{}^{'}\beta}_{i})\\
 & \text{subject to }\bm{D^{k}}\bm{\beta^{k}-\theta^{k}=0},\,\forall k\in\{1,..,H+1\}\end{eqnarray*}

Then, DP1 can be solved using the ADMM procedure similar to the one
described for DP. 

\textbf{Discussion on Benchmark Problem's complexity: }Benchmark Problem
is restated here for convenience: 

\begin{eqnarray*}
 & \textbf{Benchmark\,\ Problem (BP)} & \max_{\bm{\pi\in\Pi_{BC}}}W(R_{1}(\bm{\pi}),...,R_{N}(\bm{\pi}))\\
 &  & \text{subject to. }R_{i}(\bm{\pi})\geq R_{i}^{min},\,\forall i\in\{1,...,N\}\end{eqnarray*}

Let the power set of $\mathcal{U}$ be $S_{\mathcal{U}}$, where $S_{\mathcal{U}}$
consists of $2^{N}$ subsets of UEs. Let $S_{\mathcal{U}}(j)$ denote
the $j^{th}$ element of $S_{\mathcal{U}}$. Define a set of power
profiles,$\mathcal{P}^{S_{\mathcal{U}}}$, where the $\mathcal{P}^{S_{\mathcal{U}}}(j)$
corresponds to the $j^{th}$ element in the set and it corresponds
to the power profile when the UEs in set $S_{\mathcal{U}}(j)$ transmit
at their maximum power levels and the rest of the UEs do not transmit.
Note that for $\bm{\pi\in\Pi_{BC}}$, $\bm{\pi(t)}$ corresponds to
a power profile in $\mathcal{P}^{S_{\mathcal{U}}}$. Therefore, the
average throughput achieved by UE $i$, $R_{i}(\bm{\pi})$, where
$\bm{\pi\in}\bm{\Pi_{BC}}$, can also be expressed as $R_{i}(\bm{\pi})=\sum_{j=1}^{2^{N}}\alpha_{j}r_{i}(\mathcal{P}^{S_{\mathcal{U}}}(j))$,
with $\alpha_{j}\geq0,\forall j\in\{1,..,2^{N}\}$and $\sum_{j=1}^{2^{N}}\alpha_{j}=1$.
Here the fraction $\alpha_{j}$ associated with each profile $\mathcal{P}^{S_{\mathcal{U}}}(j)$
corresponds to the fraction of transmission time associated with that
power profile. 

Consider the following problem: \begin{eqnarray*}
 & \textbf{BP1} & \max_{\bm{\textbf{y,\ensuremath{\alpha}}}}W(y_{1},...,y_{N})\\
 &  & \text{subject to. }y_{i}\geq R_{i}^{min},\,\forall i\in\{1,...,N\}\\
 &  & y_{i}=\sum_{i=1}^{2^{N}}\alpha_{i}r_{i}(\mathcal{P}^{S_{\mathcal{U}}}(j)),\,\forall i\in\{1,...,N\}\\
 &  & \alpha_{j}\geq0,\forall j\in\{1,..,2^{N}\},\,\sum_{j=1}^{2^{N}}\alpha_{j}=1\end{eqnarray*}

Next, in order to show that the above problem is NP-hard we will show
intuitively why is it so, but the detailed proof follows from proof
of Theorem 1 in \cite{luo2008dynamic}. Consider $W(y_{1},..,y_{N})=\sum_{i=1}^{N}y_{i}$
,to be a linear function, $R_{i}^{min}=0,\,\forall i\in\{1,...,N\}$
and the cross channel gains amongst some users who do not share an
edge in the interference graph to be $0$ and the cross channel gains
amongst the interfering neighbors to be $\infty$. This implies that
in any optimal solution will correspond to the transmission by a MIS
of the interference graph. This can be justified as follows. Consider
an optimal solution in which two neighboring UEs are transmitting,
making one of the UEs not transmit will definitely increase the sum
throughput contradicting the optimality. Specifically, this problem
reduces to finding the maximum weighted maximum indpendenet set which
is NP hard. Here the weight of each MIS corresponds to $\sum_{i=1}^{N}r_{i}(\bm{p^{I_{j}}})$. 

\textbf{Proof of Theorem 1:} The success probability of Phase 1 is
high, $(1-\frac{1}{N^{c_{1}-1}})$ (lower bound), (see \cite{johansson1999simple}
for detail), here\textbf{ }we analyze Phase 2. 

We first show that, if the list of remaining colors given as, $C1_{i}^{n}$
is empty at $n\geq\lceil c_{1}\log_{\frac{4}{3}}N\rceil+\lceil c_{2}\log_{x}N\rceil+2$
and if this holds $\forall i\in\{1,...,N\}$ then the Phase 2 has
converged to a set of $H$ MISs which span all the UEs. Let us assume
otherwise, i.e. $C1_{i}^{n}$ is empty $\forall i\in\{1,...,N\}$
however, the set corresponding to some color $h\in\{1,...,H\},$ $I_{h}^{'}$
is not a MIS. $I_{h}^{'}$ has to be an IS. Assume otherwise, i.e.
$I_{h}^{'}$ is not an IS, which implies that there must exist a pair
of UEs, $i$ and $j$, which are neighbors and are a part of $I_{h}^{'}$.
If this is true then both then both acquired the color $h$ either
in the same time slot or in different time slots, in Phase 1 or 2.
In case the color is acquired in different time slots, then after
the first time slot when either of the UEs in the pair acquires the
color it will transmit the final color choice, $h$ to the neighbors
(see Table II and III) who in turn delete that color. However, if
the color is deleted by the neighbor then it cannot acquire it in
the future thus, ruling out the case that the colors were acquired
in two different time slots. If the color was acquired by the UEs
in the same time slot, then it implies that despite the conflict in
tentative choice the UEs acquire the color which is not possible (see
Table II and III). This shows that $I_{h}^{'}$ is an IS. 

Since $I_{h}^{'}$ is not maximal then $\exists$ at least one UE-$j\not\in I_{k}^{'}$
which can be added to this set without violating independence. From
the assumption, we have $C1_{j}^{n}=\phi$ which implies that the
color $h$ was deleted at some stage from the original list of all
the colors either in Phase 1 or 2. The deletion of $h$ was a result
of that color being acquired finally by at least one of the neighbors
$k\in\mathcal{N}(j)$ since $j\not\in I_{k}^{'}$. In that case, $j$
cannot acquire $h$ as it will violate the independence property.

Next, we show that indeed the list of all colors available $C1_{i}^{n}$
is empty at the end of Phase 2 with a high probability. Let $U^{n}$
correspond to the number of UEs which have a non-empty list at the
beginning of time slot $n$ and, let $Tn(U^{n})$ correspond to the
total time needed before all the UEs have an empty list. The probability
that a UE at time slot $n$ with a non-empty list will have an empty
list in next time slot is always greater than $c^{H}(1-c)^{H^{2}}$.
This can be explained as, if the UE chooses all the colors in the
list assuming (worst case $H$ number of colors remain) and all the
neighbors (worst case $H$ neighbors) do not choose any color, then
all the colors in the UE's list will be deleted. From this, we get
$E(U^{n+1})\leq(1-c^{H}(1-c)^{H^{2}})U^{n}=\frac{1}{x}U^{n}$ and
$Tn(U^{n})=1+Tn(U^{n+1})$. Assuming that the Phase 2 will start with
$N$ UEs whose list are non-empty (worst case) and from \cite{karp1994probabilistic}
we get $P(Tn(N)\geq\lceil c_{2}\log_{x}N\rceil)\leq\frac{1}{N^{c_{2}-1}}$.
This gives the lower bound on success probability of Phase 2 and thereby
the result in the Theorem. ~~~~~~~~~~~~~~~~~~~~~~~~~~~~~~~~~~~~~~~~~~~~(Q.E.D) 

\textbf{Proof of Theorem 2:} The two problems which are introduced
to transit from CP to DP are,

\begin{eqnarray*}
\textbf{Global Primal Problem\,(GPP)} & \max_{\{\beta_{i}^{k}\}_{i,k}}\sum_{k=1}^{H}W_{i}(\sum_{i=1}^{N}\beta_{i}^{k}R_{i}^{k})\\
\text{subject to} & \sum_{k=1}^{H}\beta_{i}^{k}R_{i}^{k}\geq R_{i}^{min},\sum_{k=1}^{H}\beta_{i}^{k}=1,\,\forall i\in\{1,...,N\}\\
\beta_{i}^{k}=\beta_{l}^{k},\;\forall i\not=l,\forall k\in\{1,...,H\}, & \,\,\beta_{i}^{k}\geq0,\;\forall i\in\{1,...,N\},\forall k\in\{1,...H\}\end{eqnarray*}

The second problem, Local Primal Problem (LPP) is the same as GPP
except we choose a subset of the constraints from the above problem.
Basically, instead of an equality constraint between the UE's estimate
and every other UE in the network, we only keep the equality constraints
between the UE and its neighbors, i.e. $\,\beta_{i}^{k}=\beta_{l}^{k},\,\forall k\in\{1,...,H\},\forall l\in\mathcal{N}(i)$.
This is formally stated below:

\begin{eqnarray*}
\textbf{Local Primal Problem\,(LPP)} & \max_{\{\beta_{i}^{k}\}_{i,k}}\sum_{k=1}^{H}W_{i}(\sum_{i=1}^{N}\beta_{i}^{k}R_{i}^{k})\\
\text{subject to} & \sum_{k=1}^{H}\beta_{i}^{k}R_{i}^{k}\geq R_{i}^{min},\sum_{k=1}^{H}\beta_{i}^{k}=1,\,\forall i\in\{1,...,N\}\\
\beta_{i}^{k}=\beta_{l}^{k},\;\forall l\not\in\mathcal{N}(i),\forall k\in\{1,...,H\}, & \,\,\beta_{i}^{k}\geq0,\;\forall i\in\{1,...,N\},\forall k\in\{1,...H\}\end{eqnarray*}

To show that problems CP and GPP are equivalent, we need to show that
from $\bm{\beta}^{*}=(\bm{\beta_{1}^{*},..,\beta_{N}^{*}}),$ an optimal
argument of GPP, we can obtain an optimal argument of CP, i.e. $\bm{\alpha}^{*}$
and vice versa. Since $\bm{\beta}^{*}$ is the optimal value (assuming
feasibility) we know that $\bm{\beta}_{i}^{*}=\bm{\beta}_{j}^{*}$
(component-wise) holds $\forall i,j\in\{1,...,N\}$. 

a).Let $\bm{\alpha^{'}=\beta_{i}^{*}}$. $\bm{\alpha}^{'}$ satisfies
the constraints in CP. The objective of CP at $\bm{\alpha}^{'}$ attains
the optimal value of GPP. We need to establish that $\bm{\alpha}^{'}$
is indeed the optimal argument of CP. Assume that $\bm{\alpha}^{'}$is
not the optimal value, then there exists another $\bm{\alpha}^{*}$
which is indeed the optimal. Next, using $\bm{\alpha}^{*}$, we can
obtain another $\bm{\beta}^{'}$as follows, $\bm{\beta_{1}^{'}=\alpha^{*}}$and
$\bm{\beta_{i}^{'}=\beta_{1}^{'},\,}\forall i\in\{1,...,N\}$. The
objective of GPP at $\bm{\beta}^{'}$ should be higher than $\bm{\beta}^{*}$
which contradicts $\bm{\beta}^{*}$ being the optimal argument. Note
that if either of CP or GPP is infeasible then the other problem can
be shown to be infeasible as well. On the same lines we can show that
from an $\bm{\alpha}^{*}$ we can obtain $\bm{\beta}^{*}$ as well.

b). Let $\bm{\alpha}^{*}$ be the optimal solution to CP, and define
$\bm{\beta^{''}}$a solution to GPP as follows. Let $\bm{\beta_{1}^{''}=\alpha^{*}}$
and $\bm{\beta_{i}^{''}}=\bm{\beta_{j}^{''},\,}\forall j\not=i$ and
since $\bm{\alpha}^{*}$ satisfies the constraints of CP, i.e. it
is feasible, implies that $\bm{\beta^{''}}$ as well satisfies constraints
of GPP. We want to show that $\bm{\beta^{''}}$is the optimal value
as well, assume that it is not and there exists an argument $\bm{\beta}^{*}$
for which the objective takes a higher value. If this is the case
then, from $\bm{\beta}^{*}$ we can construct a $\bm{\alpha}^{'}$as
in part a). which, if $\bm{\beta}^{*}$ takes a higher value than
$\bm{\beta^{''}}$, takes a higher value than $\bm{\alpha}^{*}$ thus,
contradicting optimality.

To show that GPP and LPP are equivalent, we use the following fact,
since LPP consists of a subset of the constraints then the solution
of LPP is an upper bound of the solution to GPP. We need to show that
the gap between the solution of LPP and GPP is always 0. Note that
for an optimal solution of LPP, $\bm{\gamma}^{*}=(\bm{\gamma_{1}^{*},..,\gamma}_{N}^{*})$
we know that $\bm{\gamma}_{i}^{*}=\bm{\gamma}_{j}^{*}\,\,\forall j\in\mathcal{N}(i)$
(component-wise). If we can show that $\bm{\gamma}_{i}^{*}=\bm{\gamma}_{j}^{*}\,\,\forall j\in\{1,...,N\}$
then LPP and GPP will be equivalent, since it will also satisfy all
the constraints of GPP. Assume that this does not hold then $\exists\, i,\, j$
such that $\bm{\gamma}_{i}^{*}\not=\bm{\gamma}_{j}^{*}$. Since, the
interference graph is connected $\exists$ a path $i\rightarrow j=\{i_{1},...,i_{s}\}$
which implies, $\bm{\gamma}_{i}^{*}=\bm{\gamma}_{i_{1}}^{*}...=\bm{\gamma}_{j}^{*}$.
This leads to a contradiction, thereby establishing the claim. 

Lastly, to show that DP is equivalent LPP. Given $\bm{\gamma}^{*}$,
define $\bm{\kappa}=\bm{\gamma^{*}}$ and a $\bm{\theta=(\theta^{1},...,\theta^{H})}$
to satisfy $\bm{D^{k}}\bm{\kappa^{k}-\theta^{k}=0},\,\forall k\in\{1,..,H\}$,
where $\bm{\kappa^{k}=}(\gamma_{1}^{*,k},..,\gamma_{N}^{*,k})$. It
can be shown using the same approach as we did for GPP and CP that
$(\bm{\kappa},\bm{\theta})$ is indeed optimal argument for DP. Assume
that $(\bm{\kappa},\bm{\theta})$ is not the optimal solution then
we know that there exists $\bm{(\kappa}^{*},\bm{\theta^{*})}$ for
which the objective in DP takes a higher value. If this is the case,
let us define $\bm{\gamma^{'}=\kappa^{*}}$, here $\bm{\gamma^{'}}$
satisfies the constraints in LPP. Also, since the objective in DP
at $\bm{(\kappa}^{*},\bm{\theta^{*})}$ takes a higher value than
that at $(\bm{\kappa},\bm{\theta})$, this yields that the objective
in LPP at $\bm{\gamma^{'}}$should take a higher value than that at
$\bm{\gamma}^{*}$, which contradicts optimality of $\bm{\gamma}^{*}$.
On the same lines, it can be easily shown that from $\bm{(\kappa}^{*},\bm{\theta^{*})}$
we can construct the optimal solution $\bm{\gamma}^{*}$ of the LPP.
This, will establish equivalence between LPP and DP. Hence, all the
four problems are equivalent. This is shown in Fig. 10. 

\begin{figure}
\centering{}\includegraphics[width=2.5in]{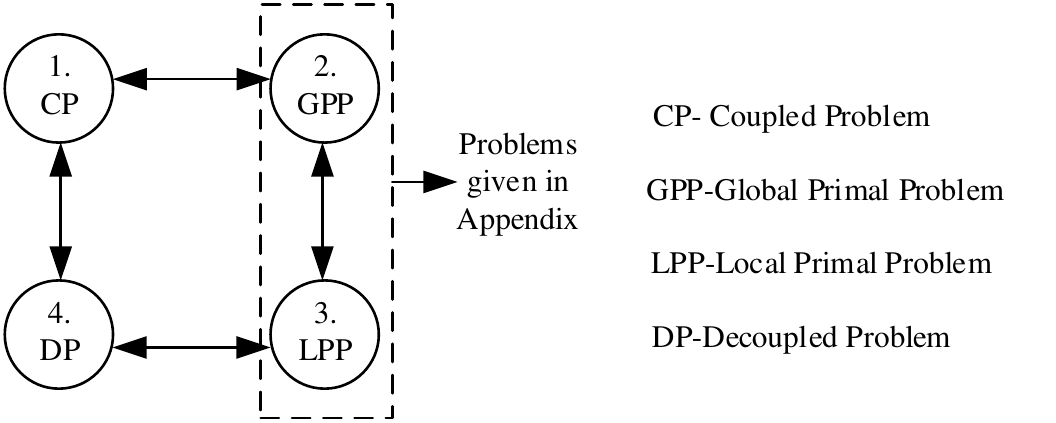}\caption{\label{fig:Problems-used-to}Problems used to transit from the Coupled
Problem (CP) to Decoupled Problem (DP).}
\end{figure}
~~~~~~~~~~~~~~~~~~~~~~~~~~~~~~~~~~~~~~~~~~~~~~~~~~~~~~~~~~~~~~~~~~~~~~~~~~~~~~~~~~~~~~~~~~~~~~~~(Q.E.D)

\textbf{Proof of Theorem 3: } According to \cite{wei20131}, the ADMM
algorithm converges with rate $O(1/P)$ if the DP is feasible and
if the feasible set is compact. Since $\mathcal{B}_{i}$ and $\Theta^{k}$
are all closed and bounded polyhedrons, the feasible set is compact.
~~~~~~~~~~~~~~~~~~~~~~~~~~~~~~~~~~~~~~~~~~~~(Q.E.D) 

\textbf{Proof of Theorem 4: }Here, we need to show three things, 

i). if $\Delta\leq\Delta^{max}$ then the distributed policy yields
a feasible solution, 

ii). the size of any MIS is $\geq\frac{N}{\Delta+1}$ , thereby using
this to show that the distributed policy, if feasible will yield a
network performance of at least $\frac{N}{\Delta+1}\log_{2}(1+\frac{p_{lb}^{max}}{(D^{ub})^{np}2^{\zeta}\sigma^{2}})$
and

iii). the upper bound on the network performance, sum throughput here
is $N\log_{2}(1+\frac{p_{ub}^{max}}{(D^{lb})^{np}\sigma^{2}})$.

i). In the Phase 1 of the algorithm the maximum number of colors used
is $\Delta+1$, since each UE selects colors from subset of $\{1,...,H\}\cap\{1,...,d_{i}+1\}$.
The first $\Delta+1$ output MISs, $\{I_{1}^{'},...,I_{\Delta+1}^{'}\}$
span all the UEs in the network. If the fraction of time assigned
to each of these $\Delta+1$ MISs is, $\alpha_{k}^{'}=\frac{R_{ub}^{min}}{\log_{2}(1+\frac{p_{lb}^{max}}{(D^{ub})^{np}2^{\zeta}\sigma_{ub}^{2}})},\,\forall k\in\{1,..,\Delta+1\}$~
then such an assignment satisfies the constraint that sum of fractions
assigned to all the colors cannot be more than 1, i.e. since $\Delta\leq\Delta^{max}\implies(\Delta+1)\frac{R_{ub}^{min}}{\log_{2}(1+\frac{p_{lb}^{max}}{(D^{ub})^{np}2^{\zeta}\sigma^{2}})}\leq1$.
Using the fact that network exhibits $\zeta-$WNI we can write the
minimum instantaneous throughput that can be obtained by UE-$i$ as,
$\log_{2}(1+\frac{p_{i}^{max}}{(D_{iT(i)})^{np}2^{\zeta}\sigma_{ub}^{2}})$,
and minimum instantaneous throughput of any UE as, $\log_{2}(1+\frac{p_{lb}^{max}}{(D^{ub})^{np}2^{\zeta}\sigma_{ub}^{2}})$.
Thus, given the fractions assigned to the MISs, $\alpha_{k}^{'}=\frac{R_{ub}^{min}}{\log_{2}(1+\frac{p_{lb}^{max}}{(D^{ub})^{np}2^{\zeta}\sigma_{ub}^{2}})},\,\forall k\in\{1,..,\Delta+1\}$,
which span all the UEs. each UE $i$'s throughput requirement is satisfied,
$,\frac{R_{ub}^{min}}{\log_{2}(1+\frac{p_{lb}^{max}}{(D^{ub})^{np}2^{\zeta}\sigma_{ub}^{2}})}\log_{2}(1+\frac{p_{i}^{max}}{(D_{iT(i)})^{np}2^{\zeta}\sigma_{ub}^{2}})\geq R_{ub}^{min}$.

ii). Assume that $\exists$ an MIS whose size is $S<\frac{N}{\Delta+1}$
. Each UE in the MIS can exclude a maximum of $\Delta$ UEs from being
included in the MIS. This implies that $S(\Delta+1)$, represents
the total number of UEs excluded and the UEs in the MIS which put
together should exceed $N$. Since this is not the case here, the
contradiction implies that $S\geq\frac{N}{\Delta+1}$. This combined
with minimum instantaneous throughput of any UE, we get the lower
bound $\frac{N}{\Delta+1}\log_{2}(1+\frac{p_{lb}^{max}}{(D^{ub})^{np}2^{\zeta}\sigma_{ub}^{2}}),$for
our policy. 

iii). The upper bound on the optimal network performance is obtained
by summing maximum instantaneous throughput of any UE $\log_{2}(1+\frac{p_{ub}^{max}}{(D^{lb})^{np}\sigma_{lb}^{2}})$
for all UEs, $N\log_{2}(1+\frac{p_{ub}^{max}}{(D^{lb})^{np}\sigma_{lb}^{2}})$.
Computing the ratio of the lower bound of proposed scheme $\frac{N}{\Delta+1}\log_{2}(1+\frac{p_{lb}^{max}}{(D^{ub})^{np}2^{\zeta}\sigma^{2}})$
and $N\log_{2}(1+\frac{p_{ub}^{max}}{(D^{lb})^{np}\sigma^{2}})$,
we get $\frac{\log_{2}(1+\frac{p_{lb}^{max}}{(D^{ub})^{np}2^{\zeta}\sigma^{2}})}{(\Delta+1)\log_{2}(1+\frac{p_{ub}^{max}}{(D^{lb})^{np}\sigma^{2}})}$
which is no less than, $\Gamma=\frac{R_{ub}^{min}}{\log_{2}(1+\frac{p_{ub}^{max}}{(D^{lb})^{np}\sigma^{2}})}$
since $\Delta\leq\Delta^{max}$.~~~~~~~~~~~~~~~~~~~~~~~~~~~~~~~~~~~~~~~~~~~~~~~~~~~~~~~~~~~~~~~~~~~~~~~~~~~~~~~~~
(Q.E.D)

\textbf{Proof of Theorem 5: }Let $\Delta^{*}=6\eta$ with $\eta=\lceil\frac{\log_{2}(1+\frac{1}{(D^{lb})^{np}\sigma_{lb}^{2}}p_{ub}^{max})}{R_{lb}^{min}}\rceil$.
We assume that the interference graph is constructed using a distance
threshold rule (Subsection V-B). Note that each UE's minimum throughput
requirement is at least $R_{lb}^{min}$, this combined with maximum
instantaneous throughput of any UE $\log_{2}(1+\frac{p_{ub}^{max}}{(D^{lb})^{np}\sigma_{lb}^{2}})$
yields that each UE needs at least $\frac{R_{lb}^{min}}{\log_{2}(1+\frac{p_{ub}^{max}}{(D^{lb})^{np}\sigma_{lb}^{2}})}$
fraction of time slots. First, we need to show that if there exists
a clique (a subset of vertices in the graph which are mutually connected)
in the interference graph of size, $X$ greater than $\eta$ then
the minimum throughput constraints cannot be satisfied. Assume that
there does exist such a clique, then any MIS based scheduling policy
will allocate separate time slots to each UE in the clique. This is
true because no two UEs in the clique will belong to the same MIS.
This implies that $X\frac{R_{lb}^{min}}{\log_{2}(1+\frac{p_{ub}^{max}}{(D^{lb})^{np}\sigma_{lb}^{2}})}$
is the total fraction separate time slots needed which has to be less
than 1. But as $X\geq\eta$, this leads to infeasibility. Next, if
$\Delta\geq$$\Delta^{*}$, we claim that we will have at least one
clique in the graph satisfying this condition. Then $\exists$ UE-$i$
with a degree $d_{i}\geq6\eta$, this implies that within a radius
of $D^{th}$ around SBS-$T(i)$ $\exists$ $6\eta$ SBSs. Also, this
circle  around SBS-$T(i)$ can be partitioned into 6 sectors subtending
$\frac{\pi}{3}$ at the center.The distance between any two points
located in the sector is $\leq D^{th}$, which we justify next. Hence,
all the points in a sector are mutually connected, thus forming a
clique.

Let the 2-D polar coordinates of two points $i,j$ in a sector be
$(r_{i},0)$ and $(r_{j},\theta)$, where $0\leq r_{i}\leq D^{th},0\leq r_{j}\leq D^{th}\,$and
$0\leq\theta\leq\frac{\pi}{3}$. Hence, the square of the distance
between the two points is expressed as $f(r_{i},r_{j},\theta)=r_{i}^{2}+r_{j}^{2}-2r_{i}r_{j}cos\theta$
and our claim is that the maximum value $f(r_{i},r_{j},\theta)$,
in the set of constraints above is no greater than $(D^{th})^{2}$.
We formally state this as an optimization problem below: \begin{eqnarray*}
\max_{r_{i},r_{j},\theta} & f(r_{i},r_{j},\theta)\\
 & 0\leq r_{i}\leq D^{th},0\leq r_{j}\leq D^{th}\,\\
 & 0\leq\theta\leq\frac{\pi}{3}\end{eqnarray*}

Since, both $r_{i},r_{j}$ are non-negative, this implies that in
the above optimization problem, $\theta=\frac{\pi}{3}$ has to be
satisfied in the optimal argument. Substituting $\theta=\frac{\pi}{3}$
in $f(r_{i},r_{j},\theta)$ we get, $f(r_{i},r_{j},\frac{\pi}{3})=r_{i}^{2}+r_{j}^{2}-r_{i}r_{j}$.
Next, we show that $r_{i}^{2}+r_{j}^{2}-r_{i}r_{j}\leq(D^{th})^{2}$
for $0\leq r_{i}\leq D^{th},0\leq r_{j}\leq D^{th}$. Fix a $0\leq r_{j}\leq D^{th}$,then
$r_{i}^{2}+r_{j}^{2}-r_{i}r_{j}$ takes its maximum value at $r_{i}=D^{th}$,
which gives $(D^{th})^{2}+r_{j}^{2}-D^{th}r_{j}$. Since $0\leq r_{j}\leq D^{th}$,
this yields $(D^{th})^{2}+r_{j}^{2}-D^{th}r_{j}\leq(D^{th})^{2}$
which establishes the claim. 

If we have a total of $6\eta$ SBSs in the circle then at least one
sector has to have more than $\eta$ SBSs (Pigeonhole principle),
which implies that a clique of size $X\geq\eta$ will exist. ~~~~~~~~~~~(Q.E.D)

%
{}

\bibliographystyle{IEEEtran}
\bibliography{Distbtd_Femto_jmtd}

\end{document}